\documentclass{pasa}%

\title[Nearby Lyman alpha galaxies]{Lyman alpha Emitting Galaxies in the Nearby
Universe}

\author[M.Hayes]{Matthew Hayes\\ 
\affil{Department of Astronomy, The Oskar Klein Centre for Cosmoparticle
Physics, Stockholm University, AlbaNova University Centre, SE-106 91 Stockholm,
Sweden}
\affil{$\star$\,\href{mailto:matthew@astro.su.se}{matthew@astro.su.se}}}%

\jid{PASA}
\doi{10.1017/pas.\the\year.2014}
\jyear{\the\year}

\usepackage[authoryear]{natbib}
\usepackage{graphicx}
\usepackage{floatrow}
\usepackage{caption}
\usepackage{subcaption}
\usepackage{amsmath}
\usepackage{amssymb}
\usepackage{aas_macros}
\usepackage[hyperfootnotes=false]{hyperref}
\hypersetup{
	colorlinks=false,
	citecolor=Violet,
	linkcolor=Red,
	urlcolor=Blue}
\usepackage[usenames,dvipsnames]{xcolor}

\newcommand{\lya}{\mbox{Ly$\alpha$}}

\newcommand{\halpha}{\mbox{H$\alpha$}}
\newcommand{\hbeta}{\mbox{H$\beta$}}

\newcommand{\hI}{\mbox{H{\sc i}}}
\newcommand{\hII}{\mbox{H{\sc ii}}}
\newcommand{\oI}{\mbox{O{\sc i}}}
\newcommand{\oII}{\mbox{O{\sc ii}}}
\newcommand{\oIII}{\mbox{O{\sc iii}}}

\newcommand{\cII}{\mbox{C{\sc ii}}}
\newcommand{\sII}{\mbox{S{\sc ii}}}
\newcommand{\sIII}{\mbox{S{\sc iii}}}
\newcommand{\siII}{\mbox{Si{\sc ii}}}

\newcommand{\nII}{\mbox{N{\sc ii}}}

\newcommand{\kms}{\mbox{km\,s$^{-1}$}}
\newcommand{\ergsec}{\mbox{erg~s$^{-1}$}}

\newcommand{\ergseccm}{\mbox{\mbox{erg~s$^{-1}$~cm$^{-2}$}}}
\newcommand{\msun}{\mbox{M$_\odot$}}

\newcommand{\msunyr}{\mbox{M$_\odot$~yr$^{-1}$}}
\newcommand{\fesclya}{\mbox{$f_\mathrm{esc}^{\mathrm{Ly}\alpha}$}}
\newcommand{\fc}{\mbox{$f_\mathrm{c}$}}
\newcommand{\ebv}{\mbox{$E_{B-V}$}}

\newcommand{\percm}{\mbox{cm$^{-2}$}}
\newcommand{\arcsec}{\mbox{$^{\prime\prime}$}}

\newcommand{\mpccube}{\mbox{Mpc$^3$}}

\newcommand{\lstar}{\mbox{$L^\star$}}

\newcommand{\mhi}{\mbox{$M_\mathrm{HI}$}}
\newcommand{\pcyg}{\mbox{P\,Cygni}}
\newcommand{\nhi}{\mbox{$N_\mathrm{H{\textsc i}}$}}

\newcommand{\ewlya}{\mbox{$W_{\mathrm{Ly}\alpha}$}}
\newcommand{\ewha}{\mbox{$W_{\mathrm{H}\alpha}$}}
\newcommand{\xilya}{\mbox{$\xi_{\mathrm{Ly}\alpha}$}}

\newcommand{\iras}{\mbox{IRAS\,08339+6517}}
\newcommand{\izw}{\mbox{{\sc i}\,Zw\,18}}
\newcommand{\sbs}{\mbox{SBS\,0335--052}}
\newcommand{\har}{\mbox{Haro\,11}}
\newcommand{\eso}{\mbox{ESO\,338--IG04}}

\begin{document}%
\begin{abstract}
The Lyman alpha emission line (\lya) of neutral hydrogen (\hI) is intrinsically
the brightest emission feature in the spectrum of astrophysical nebulae, making
it a very attractive observational feature with which to survey galaxies.
Moreover as an ultraviolet (UV) resonance line, \lya\ possesses several unique
characteristics that make it useful to study the properties of the interstellar
medium (ISM) and ionizing stellar population at all cosmic epochs.  In this
review I present a summary of \lya\ observations of galaxies in the nearby
universe.  By UV continuum selection, at the magnitudes reachable with current
facilities, only $\approx 5$\% of the local galaxy population shows a \lya\
equivalent width (\ewlya) that exceeds 20~\AA.  This fraction increases
dramatically at higher redshifts, but only in the local universe can we study
galaxies in detail and assemble unprecedented multi-wavelength datasets.  I
discuss many local \lya\ observations, showing that when galaxies show net \lya\
emission, they ubiquitously also produce large-scale halos of scattered \lya,
that dominate the integrated luminosity.  Concerning global measurements, we
discuss how \ewlya\ and the \lya\ escape fraction (\fesclya) are higher
(\ewlya~$\gtrsim 20$\AA\ and \fesclya~$\gtrsim 10$\%) in galaxies that represent
the less massive and younger end of the distribution for local objects.  This is
connected with various properties, such that \lya-emitting galaxies have lower
metal abundances (median value of $12+\log(\mathrm{O/H})\sim 8.1$) and dust
reddening.  However, the presence of galactic outflows/winds is also vital to
Doppler shift the \lya\ line out of resonance with the atomic gas, as high
\ewlya\ is found only among galaxies with winds faster than $\sim 50$\kms.  The
empirical evidence is then assembled into a coherent picture, and the
requirement for star formation driven feedback is discussed in the context of an
evolutionary sequence where the ISM is accelerated and/or subject to
hydrodynamical instabilities, which reduce the scattering of \lya.  Concluding
remarks take the form of perspectives upon future developments, and the most
pressing questions that can be answered by observation.
\end{abstract}
\maketitle%

\section{INTRODUCTION}

The Lyman alpha (\lya) emission line of atomic hydrogen (\hI) is intrinsically
the most luminous spectral emission feature in astrophysical nebulae.  It is
produced by the spontaneous decay from the first excited state to the ground
state ($n=2\rightarrow1$, where $n$ is the principle quantum number), where the
energy difference between the levels dictates a photon energy of 10.2~eV, or
wavelength $\lambda=1215.67$~\AA.  After the capture of an electron in ionized
gas, the transition probabilities of the following radiative cascade are such
that 68\% of recombinations involve the production of a \lya\ photon
\citep{Dijkstra2014}.  Thus if all ionizing photons had an energy slightly above
the ionization edge of \hI, $\approx 50$\% of the total ionizing energy would be
reprocessed into the \lya\ line.  Because the line may be intrinsically so
luminous, \lya\ was initially proposed as the spectral beacon by which to
identify the first generations of primeval galaxies almost five decades ago
\citep{Partridge1967}.

\lya\ plays a pivotal role in contemporary astrophysics, where it is used to
identify high-redshift ($z$) star-forming galaxies by either narrowband filter
observations or spectroscopic techniques, and frequently to confirm the redshift
of candidate galaxies selected by other methods.  Indeed the recognition that
\lya\ observations have attained makes the detection of \lya\ a primary science
goal for many new instruments on large telescopes: first light has recently been
seen by HyperSuprime Cam \citep[HSC,][]{Takada2010} at the Subaru Telescope, the
Multi Unit Spectroscopic Explorer \citep[MUSE,][]{Bacon2010} at ESO's Very Large
Telescope, and the Cosmic Web Imager \citep[CWI,][]{Martin2010} at Palomar.
Furthermore, the Hobby Eberly Dark Energy EXperiment \citep[HETDEX,][]{Hill2008}
will find on the order of $10^6$ \lya\ galaxies, and \lya\ detection is among
the science goals for spectroscopic instruments on the James Webb Space
Telescope (JWST) and all plans for Extremely Large Telescopes (ELTs). 

The above introductory paragraphs on \lya\ production are incomplete.  While
\lya\ forms in astrophysical nebulae at the stated intensities, the transition
is a resonant one, and \lya\ is also absorbed by \hI\ in the same transition.
After absorption to the $^2P$ level (in the absence of electron collisions),
there is is no alternative but for the electron to de-excite through \lya.  The
optical depth of \hI, as seen in the core of the absorption line, is given by
$\tau_0 \approx 3\times 10^{-14} \sqrt{10^4 / T}\cdot N_\mathrm{HI}$, where $T$
is the temperature of the gas and \nhi\ the column density in \percm\
\citep{Verhamme2006}.  Thus at the limiting temperatures to which hydrogen can
remain neutral, \hI\ becomes optically thick to \lya\ at $N_\mathrm{HI}\approx 3
\times 10^{13}$~\percm.  Assuming a number density of 1 atom per cubic cm, a
cloud will exceed $\tau=1$ when its diameter exceeds $10^{-5}$~pc, or just 2 AU.
Taking the Milky Way as an example, there are very few sightlines through which
\nhi\ drops below $10^{20}$~\percm\ \citep{Kalberla2005}, implying that \lya\
would almost always see upwards of $10^6$ optical depths.

The upshot is that in most galaxies, \lya\ undergoes a radiative transfer
process: photons scatter until they either escape from the galaxy or are
absorbed by a dust grain, and dust extinction is also strongest in the far UV.
This transfer may be thought of as a diffusion-like process, where photons take
random walks in both physical and frequency space \citep{Osterbrock1962}.  The
path taken by \lya\ is entirely regulated by the distribution of \hI\ that it
encounters and must traverse, which in turn determines the likelihood that \lya\
will encounter dust.  Fortunately \lya\ may see a significantly lower optical
depth if it is shifted in frequency or the \hI\ is moving; the former can occur
either after many scattering events as it diffuses in frequency through the
redistribution profile, or by scattering in \hI\ that is itself kinematically
offset from the \hII\ media where the \lya\ formed.  Ultimately the emitted
\lya\ luminosity (also its EW and departure from intrinsic \lya/\halpha\ ratio
of 8.7) will be a function of \hI\ distribution, gas kinematics, dust content,
and galaxy viewing angle.

\subsection{Key Applications of Lyman alpha}\label{sect:intro:key}

\lya\ transfer makes it hard to interpret the observed \lya\ flux and EW from an
individual galaxy, because the escape fraction, \fesclya, is difficult to
predict for given configuration.  However the transfer process, and the
sensitivity of \lya\ to different ISM properties, is also one of the major
advantages of the transition.  I now outline some key applications.

$\bullet$ \emph{The Evolution of Galaxies.} The fraction of galaxies with
\ewlya~$\gtrsim 20$\AA\ (the canonical definition of a \lya-emitter, LAE), at
absolute UV magnitudes brighter than --18, is just 5\% in the nearby universe
\citep{Cowie2010}, where LAEs are rare.  However this fraction increases
strongly with increasing redshift, to $\sim25$\% at $z \approx 3$
\citep{Shapley2003} to over 50\% at $z=6-6.5$ \citep{Stark2010,Curtis-Lake2012}.
\fesclya\ evolves even more strongly over the same redshift range
\citep{Hayes2011evol,Blanc2011}.  This monotonic evolution, that spans a factor
of 100 in \fesclya, is a key result of many \lya\ surveys but has no conclusive
explanation.  Dust and \hI\ covering have both been suggested, and the answer
must indeed lie among the quantities mentioned above, or combinations thereof,
in the co-evolving properties of stars, gas and dust.

$\bullet$ \emph{The Epoch of Reionization.} \lya\ emission offers a unique
opportunity to study the ionization state of intergalactic medium (IGM) at an
epoch where other methods -- e.g. the Gunn-Peterson trough in QSO spectra and
Thompson scattering of the cosmic microwave background (CMB) -- are insensitive.
The \lya\ emitter fraction and \fesclya\ evolution discussed above reverses
after $z\sim 6.5$, decreasing to $\lesssim 20$\% \citep[e.g.][]{Pentericci2014}
at $z\sim 7$.  Possible interpretations include an increase in the ionizing
photon escape fraction \citep[][which is anyway needed for
reionization]{Dijkstra2013}, but also that and increasingly neutral IGM starts
to absorb the \lya\ produced by the galaxies themselves.  Disentangling the
scenarios requires more information and solid constraints on the reionization
history and topology require much larger samples \citep[e.g.][]{Jensen2014}, but
these will become available in the coming years with HSC.  Moreover, as the IGM
becomes neutral, the damping wing of \hI\ \lya\ absorption may begin to affect
the profile shape of the \lya\ line that is transmitted, giving \lya\ another
unique application.

$\bullet$ \emph{Galaxy Kinematics.} \lya\ scatters coherently in the restframe
of the \hI\ atom, and at scattering events is shifted in frequency by the
velocity of the scattering medium.  Thus as \lya\ may escape from galaxies
because of frequency shifts, the kinematic structure of the atomic gas becomes
imprinted onto the line.  This manifests as both a redshift (for outflowing gas)
of the centroid of the main emission peak \citep[e.g.][]{Hashimoto2013} and also
as characteristic features that modify the shape of the line profile
\citep[e.g.][]{Verhamme2008}.  While there are many probes of kinematics in
astrophysics, nebular line kinematics exclusively traces the warm ionized
medium.  \lya\ kinematics on the other hand is shaped by kinematic differences
between \hII\ (production) and \hI\ (scattering) media.  Moreover \lya\ is
intrinsically very bright, and can be seen redshifted from the most distant
galaxies.  This again provides unique insights into the evolution of the ISM of
galaxies, which can only feasibly be done with \lya.

$\bullet$ \emph{Atomic Gas Surrounding Galaxies.}  As well as modifying the line
profile, scattering also changes the surface brightness profile of emitted \lya\
\citep{Steidel2011,Hayes2013}.  The mechanisms by which galaxies obtain the gas
they need to fuel star formation is one of the most pressing issues in
extragalactic science \citep[e.g.][]{Keres2005,Dekel2009}, and necessitates a
knowledge of the \hI\ distribution outside of star-forming regions and into the
circumgalactic medium (CGM).  Probing this circumgalactic \hI\ is
observationally very challenging.  In principle it can be done by
$\lambda=21$~cm observations of \hI\ directly, but current telescopes cannot
push such techniques beyond the very local universe.  An alternative is to use
absorption spectroscopy of background QSOs that pierce galaxy halos at different
impact factors, enabling us to measure \hI\ temperatures, densities and
kinematics \citep[e.g.][]{Lanzetta1995,Tumlinson2013,Danforth2014}.
Unfortunately appropriately bright QSOs are rare and thus studies, while rich
with information,  are limited to statistical studies of the average galaxy.  A
promising third method is to illuminate the circumgalactic \hI\ with \lya\
produced in the central star-forming regions.  Indeed \lya\ is perfect for such
an application, being both the brightest intrinsic emission line, and being
resonant in precisely the medium we need in order to image the CGM.

\subsection{Lyman alpha Observations of the Nearby Universe -- This Review}

Section~\ref{sect:intro:key} presents the main astrophysical applications of
\lya\ emission, both as a diagnostic of the galaxies themselves and the IGM.
The key difficulties of observing high-$z$ galaxies are that fluxes are low and
high signal-to-noise data are hard to obtain, that spatial information is
minimal or absent, and that important features are redshifted away from
atmospheric transmission windows.  In the local universe surface brightness is
higher by a factor of $(1+z)^3$, and spatial sampling can become almost
arbitrarily high.  Moreover, only in the local universe can we assemble the
complete set of multi-wavelength observations, including but not limited to, all
the continuum bands that probe both hot and cold stars, emission lines that
provide a wealth of intrinsic diagnostics of the nebulae in which \lya\ forms,
direct measurements of far infrared continuum for both hot and cold dust, direct
\hI\ measurements at $\lambda=21$~cm, X-ray observations of coronal gas, and
many more. 

Indeed the science objectives discussed in Section~\ref{sect:intro:key} can,
with the exception of reionization, all be undertaken in the low-$z$ universe.
Here \lya\ provides a unique suite of information about the ISM of galaxies that
still cannot be extracted using other techniques.  This makes \lya\ an import
observable to obtain in any thorough study of (particularly star-forming) local
galaxies.  Moreover the question can be inverted: when as much information on
the dust and gas content (distribution, kinematics, etc) has been assembled,
observations of \lya\ can then be used to calibrate our understanding of the
\lya\ transport mechanisms, and the effects of dust, gas, star-formation
evolutionary stage may all be disentangled.  This is the way, for example, we
will assemble the relevant knowledge to interpret the evolution of the \lya\
fraction with redshift.  In turn we will be able to calibrate \lya\ for high-$z$
galaxy surveys, which will soon deliver $\sim 1$ million objects, by using local
\lya\ emitters as analogues -- laboratories in which to dissect in detail the
processes ongoing in high-$z$ systems.  Local \lya\ observations will allow us
to read kinematic information off the line profile and conversely to predict the
flux, EW, and line profile shapes, precisely as needed to address topics such as
cosmic reionization that really hinge upon knowing the spectral profile.  Indeed
this is one of the major legacies established by our ultraviolet satellites; the
only difficulties are that such satellites are both expensive and competitive.   

This review focuses mainly upon empirical studies of \lya\ emission and
absorption in star-forming galaxies in the local universe.  Somewhat
arbitrarily, I have defined the `local universe' to mean redshifts where
space-based facilities are needed to observe \lya.   In principle this means
$z\lesssim 1.7$ or so, but the most distant samples discussed are at $z\sim 1$,
and thus we are considering roughly the latter half of cosmic time.  Where
appropriate I may concentrate upon what about galaxies teach us about \lya, or
about what \lya\ teaches us about galaxies.  The layout of the remainder is as
follows: \\ 
$\bullet$ In Section~\ref{sect:prehistory} I present a brief history of \lya\
observations in the local universe, which were ongoing at a time when the first
generations of high-$z$ searches were also beginning.  This concerns the first
vacuum UV observations of active galactic nuclei (AGN) and star-forming galaxies
using low dispersion spectrographs. \\ 
$\bullet$ In Section~\ref{sect:hstspec} we discuss how the \emph{Hubble Space
Telescope} changed the landscape by providing high-resolution spectra that can
resolve the \lya\ feature and interstellar absorption lines, thereby probing
atomic gas kinematics and covering. \\
$\bullet$ Section~\ref{sect:hstimage} is concerned with \lya\ imaging
observations, also from HST, that simultaneously resolve very fine structures
and reveal large-scale, diffuse \lya\ halos. \\ 
$\bullet$ In Section~\ref{sect:galex} I present a large number of key results
from survey data, that aim to answer questions about how various globally
measured properties influence \lya\ emission and under what conditions \lya\ can
be expected to be bright.  \\ These
Sections~\ref{sect:prehistory}--\ref{sect:galex} aim to establish empirically
how we have arrived at the current state-of-the-art. \\ 
$\bullet$ In Section~\ref{sect:govprop} I then synthesize all the observational
data from the previous Sections, and introduce some more speculative discussion
about how various processes fit together. \\ 
$\bullet$ Finally I do not present explicit conclusions, but close the review in
Section~\ref{sect:conc} with a number of perspectives and pressing open
questions.  These are concerned future observations and uses of \lya\ at both
low- and intermediate-$z$, with a view to understanding galaxy formation.

\section{LYMAN ALPHA OBSERVABLES}\label{sect:strength}

The literature makes use of several observable quantities that pertain to \lya,
that often are specific to how much escapes or aim to illustrate how `strong'
the feature is.  This Section summarizes some of these quantities and
conventions, and lists several caveats that may be considered while reading.

\subsection{Flux, Luminosity, and Equivalent Width}\label{sect:observable:stars}

\lya\ is mainly produced by recombinations in photoionized nebulae, where under
standard Case B assumptions 68\% of ionizing photons absorbed by hydrogen are
reprocessed into \lya\ in the following radiative cascade (See
\citealt{Dijkstra2014} for an intuitive explanation).  For continuously
star-forming galaxies, with constant star formation rate (SFR), the \lya\
equivalent width (EW) ranges between $\approx 80$\AA.  For very young systems
the EW peaks around 300\AA\ \citep{Charlot1993}, while very high EWs that exceed
1000~\AA\ may in principle be expected for very low metallicity, population {\sc
iii} stellar systems \citep{Schaerer2003,Raiter2010}.  Naturally if the SFR is
declining the intrinsic \lya\ EW may take any value less than these, and it is
worth noting that for a simple stellar population (SSP), \ewlya\ exceeds 20~\AA\
only during the first 6~Myr \citep{Leitherer1999}.

Measuring the flux and EW of most emission lines is straightforward.  However
for \lya\ this is not necessarily so and these quantities may depend upon both
methodology and definition.  As a resonant transition, both nebular \lya\ and
continuum radiation with wavelength $\lambda = 1216$~\AA\ may be absorbed by
\hI.  Depending upon the column density, this absorption may reach equivalent
widths of several tens of \AA, which is a substantial fraction of the nebular
flux.  Fluxes measured in a given aperture may or may not be reduced by this
amount.  Narrowband imaging, for example, needs to be continuum-subtracted and
therefore measures the sum of nebular emission and absorption.  Spectroscopic
observations, on the other hand, may enable the observer to isolate the
components and separate nebular emission from ISM absorption, should this be the
goal of the measurement.  However even with in spectroscopic mode isolating the
emission-only flux depends upon the spectral resolution, and the separation will
be much easier with high-resolution slit spectrographs than low-resolution
survey telescopes.

Not only is \lya\ absorbed in the ISM but also, depending on temperature and the
properties of their winds, in the atmospheres of stars.  For the hottest O stars
\lya\ EWs may be negligible, but as the population ages or the star formation
rate declines, the dominant source of UV continuum will shift to progressively
cooler stars.  \citet{Valls-Gabaud1993} showed that \lya\ measurements from some
local galaxies may be subject to significant uncertainties from stellar
absorption, and recent models by \citet{Pena-Guerrero2013} demonstrate that
stellar \lya\ absorption may reach EWs of $-30$~\AA.  For example, the effect of
stellar absorption may also vary from O-star dominated systems where the nebular
EW is high and the stellar feature is negligible, to later B-star systems where
the reverse is true.  The stellar feature may therefore shorten the timescale
over which an episode of star-formation remains bright in \lya, and the effect
may even be seen on resolved scales within a galaxy.

\subsection{\lya/Balmer Ratios and Escape Fraction}

Equivalent widths have the advantage that only a short range in wavelength is
needed to make the measurement, over which the effects of interstellar dust
(reddening curve, total extinction) should have a negligible effect.  As
discussed above, the evolutionary phase of star formation dominates the
intrinsic EW.  \lya/Balmer line ratios, however, may also be used as a measure
of the strength of \lya, and because the intrinsic ratios are limited to a
narrow range of values, have other advantages.  For example the \halpha\ line
($\lambda=6564.61$\AA) is a well-known, calibrated tracer of current
star-formation activity in nearby galaxies \citep[e.g.][]{Kennicutt1983rate}.
For Case B nebulae at temperatures in the range 5,000--20,000~K and electron
density in the range $n_\mathrm{e}=10^2$--$=10^4$~cm$^{-3}$, the \lya/\halpha\
ratio ranges between 8.1 and 11.6 \citep[][]{Hummer1987}.  Thus deviations from
the intrinsic line ratios encode information about the \hI\ scattering and dust
absorption.  While different authors do tend to adopt slightly different values,
the range of permitted values is relatively narrow.  For this review we will
adopt the value of \lya/\halpha$=8.7$, which for $T=10^4$~K gas corresponds to
$n_e\approx 350$~cm$^{-3}$, and as a convention can be traced back to at least
\citet{Hu1998}.

Frequently we make reference to the \lya\ escape fraction (\fesclya), in an
effort to find a quantity that most closely reflects the combined impact that
gas and dust have on suppressing the emission line.  We define \fesclya\ as the
ratio of the emitted \lya\ luminosity to that intrinsically produced, but
naturally a number of assumptions can enter our estimates of the intrinsic
value.  The most robust estimates of \fesclya\ will naturally come from
comparing \lya\ with other hydrogen recombination lines, where in practice
\halpha\ is the strongest and easiest to observe.  Assuming that \halpha\ is
unobscured, \fesclya\ will simply be the observed \lya/\halpha\ ratio divided by
its intrinsic case B value (8.7 as mentioned above).  Of course \halpha\ can
also be significantly absorbed by dust, and in local `normal' galaxies suffers
about 1 magnitude of extinction on average \citep{Kennicutt1983halpha}.
Obviously redder hydrogen lines would be better as they suffer less extinction
but also become systematically weaker in flux, and become harder to observe in
the near infrared.  Radio recombination lines would be optimal, suffering no
extinction at all, but are even more challenging to observe beyond the very
local universe.  Thus often the best route to \fesclya\ is to dust-correct
\halpha\ using the \hbeta\ line, which should recover all the star formation
down to moderate optical depths \citep{Hayes2005,Atek2009galex}.  However when
\halpha\ becomes very optically thick, in very dusty star-forming galaxies
\citep[e.g.][]{Martin2015}, even dust-corrected \halpha\ traces only a small
fraction of the true ionized gas, making the inferred escape fraction prone to
strong biases.  In such instances, one may do better by comparing the
\lya-derived SFR with that estimated from dust emission in the FIR, under the
assumption that systematic errors on the SFR calibrations are smaller than the
fraction of \halpha\ that is recoverable.  In the highest optical depth regimes
this is probably true.

\section{THE FIRST LOW REDSHIFT LYMAN ALPHA OBSERVATIONS}\label{sect:prehistory}

The major contributing observatories in this field are just three: the
\emph{International Ultraviolet Explorer} (IUE), the \emph{Hubble Space
Telescope} (HST), and the \emph{Galaxy Evolution Explorer} (GALEX), while
piecemeal observations have been contributed by other telescopes.  Restframe FUV
observations began with the IUE in 1977, HST started operation just 13 years
later and is still returning \lya\ data at $z<1$ some 25 years on; the
demarcation between history and present is arbitrary of course.  For the sake of
this review, I will adopt the pre-HST era, which is almost exclusively the IUE
and thus is also restricted by method to spectrophotometry of pre-selected
targets.  In the `modern' era we then have two main operational tools.  Firstly
HST, which like IUE performs pointed observations of individual targets, but in
doing so yields data that are always rich with features since both spatial and
spectral resolutions are significantly higher.  Secondly, the GALEX satellite is
similar to the IUE in resolving power but its strength comes instead from its
1.2 degree field of view, which provided the survey efficiency to yield the
statistical significance that was not available at low-$z$ with any other
telescope. 

\subsection{The First \lya\ Spectra of Active Galactic Nuclei}

Even before IUE some extragalactic objects were observed with sounding rocket
experiments, which were able to launch small UV telescopes above enough of the
atmosphere to observe in the UV.  Specifically targeting the first known quasar
3C\,273, rocket payloads provided the first measurement of \lya\ and Balmer
emission lines from any astrophysical body \citep{Davidsen1977}.  Combining the
measured \lya\ flux with optical measurements revealed a \lya/\hbeta\ ratio of
4, a measurement that then sat in stark contrast to the value of $\approx 40$
that was expected for nebular recombinations and a \lya\ enhancement from
collisions.  In 1977 Davidsen et al. discussed the order-of-magnitude \lya\
deficit with the following statement: \begin{quotation} \emph{The most obvious
explanation, that the ultraviolet lines are attenuated by absorption by dust
similar to that observed in the interstellar medium, seems untenable in view of
observations of Paschen $\alpha$ that indicate the hydrogen lines are
unreddened. But, dust which is distributed within the line-emitting gas might
destroy L$\alpha$ without having much affect on the Balmer and Paschen lines if
the nebula has high optical depth to L$\alpha$ photons. \\ \textup{[\,\dots]} \\
Whatever mechanism is at work, an understanding of the reduced \lya/\hbeta\
ratio may lead to a vastly improved knowledge of the physical conditions in QSO
envelopes.} \end{quotation} \noindent While this review is not focused upon
QSOs, replacing ``QSO" with ``galaxy" in this statement comes precisely to the
point.  Today \lya\ observations play a major role in understanding the
interstellar and circumgalactic media of galaxies at effectively all redshifts. 

The 1978 launch of IUE opened up the restframe UV to systematic study.  IUE had
a single 20\arcsec$\times$10\arcsec\ entrance aperture and \emph{Short} and
\emph{Long-Wavelength Prime} channels (SWP and LWP, respectively) that could
provide $R\approx 250$ spectroscopy between 1150 and 3000 \AA.  First
observations were again turned to QSOs and various Seyfert galaxies, and
immediately showed \lya\ to be weaker than expected for recombination theory,
with \lya/\hbeta\ and \halpha/\hbeta\ values almost never falling along
characteristic reddening curves, and \lya\ always being preferentially
suppressed  \citep{Oke1979,Wu1980,Lacy1982}.  Interpretations varied, with
suggestions that multiple scatterings of \lya\ could be responsible, that
broad-line regions experience a wide variation in their extinction laws,
ionization/excitation from already-excited states, and that there may be no
representative intrinsic spectral shape for Seyfert galaxies \citep{Wu1980}.
Indeed \citet{Lacy1982} concluded that the low observed \lya\ fluxes were likely
the result of a combination of reddening, high densities, and high \hI\ optical
depth.  All these effects conspire in the same direction but as no single
dominant quantities could be identified, it started to become clear that \lya\
transfer in true astrophysical objects is a complicated multi-parametric
process.

\subsection{Star-forming Galaxies} 

When IUE was first turned to star-forming galaxies, the results directly
mirrored those obtained in both nearby QSOs and also those being reported from
high-$z$ blind narrowband and spectroscopic surveys \citep[see][for a review,
and the forthcoming PASA review in this series by S. Malhotra]{Pritchet1994}:
\lya\ was either absent or unexpectedly weak in all galaxies. 

\begin{figure}[t!]
\begin{center}
\includegraphics[width=0.9\textwidth, angle=0]{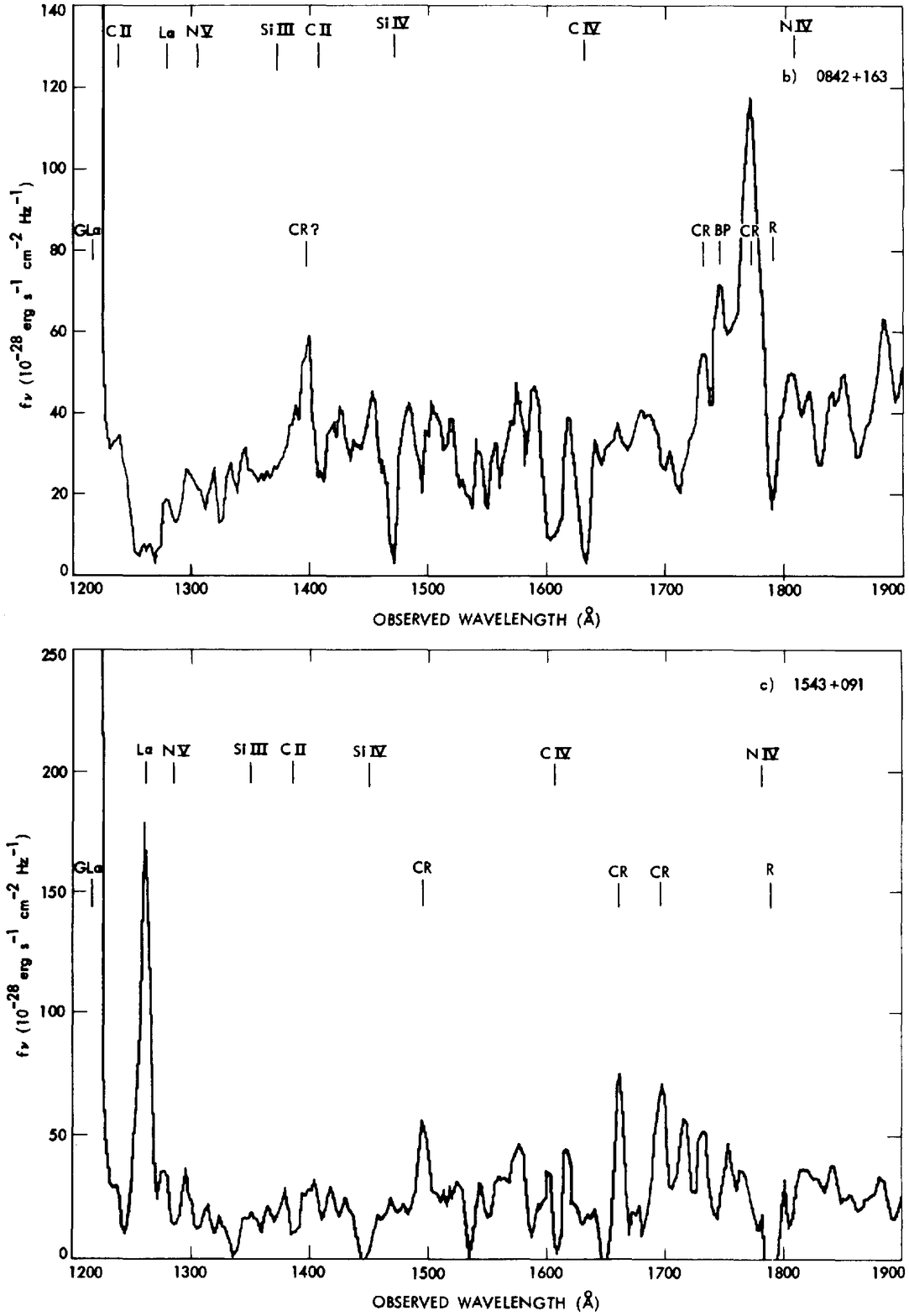}
\caption{The first \lya\ spectra of \hII\ galaxies.  Main transitions and cosmic
ray hits are labelled. The \emph{upper} panel shows Mrk\,701 (C\,0842+163), with
a redshift of $z=0.0522$ and a metallicity of $Z\approx 0.4 Z_\odot$.  At the
wavelength of $\lambda=1278$\AA, the galaxy shows only a \lya\ absorption
feature.  The \emph{lower} panel shows a very different object -- C\,1543+0907,
with a redshift of $z=0.0366$ and metallicity of 1/10 the solar value.  At the
wavelength of $\lambda=1260$\AA, the bright high-contrast \lya\ emission line is
obvious.  \emph{Reproduced by permission of the AAS, from
\citet{Meier1981}}.}\label{fig:mt81}
\end{center} 
\end{figure}

With the intent of studying the analogues of primeval galaxies at high-$z$,
\citet{Meier1981} found \lya\ in emission from just one of three nearby
\hII-selected galaxies, and in that single case with a flux well below that
expected for the nebular dust content (Figure~\ref{fig:mt81}).  This led them to
conclude that under normal conditions, \lya\ emission would be an unlikely
phenomenon.  It was determined that if a normal prescription for dust
attenuation were used to explain the \lya/\hbeta\ ratio, this would greatly
over-predict the \halpha/\hbeta\ ratio compared to observation.  Similar
conclusions were reached by \citet{Hartmann1984}, who furthered the discussion,
showing how a mixed medium of \hI\ and dust could preferentially suppress \lya,
and that static \hI\ columns of density above $10^{19}$~\percm\ would be needed
to reconcile the line ratios.  Further, they raised the issue that, where \hI\
21~cm data are available, most blue compact galaxies (BCGs) are found to sit
inside extended \hI\ halos of sufficient column density to reproduce the
measured fluxes.  All signs pointed towards the fact \lya\ emission would not be
the good observational beacon to identify primeval galaxies in the early
universe that \citet{Partridge1967} had predicted.

While the influence of \hI\ on \lya\ visibility was starting to be seen
empirically, some correlation with the dust abundance should still be expected,
albeit with a large spread.  After subsequent data acquisition, the
anticorrelation between \lya\ EW and gas-phase metallicity ($Z$) was discovered
\citep{Hartmann1988,Calzetti1992,Terlevich1993,Charlot1993}, seemingly
confirming the prediction.  However the spread on the relation was worryingly
large, and it was clear that something beyond variations in the extinction law
\citep[e.g.][]{Valls-Gabaud1993} were behind the wide range of line ratios, with
ISM geometry and holes being the most often cited scenarios.  

\begin{figure*}[t!]
\begin{center}
\includegraphics[width=0.75\textwidth, angle=0]{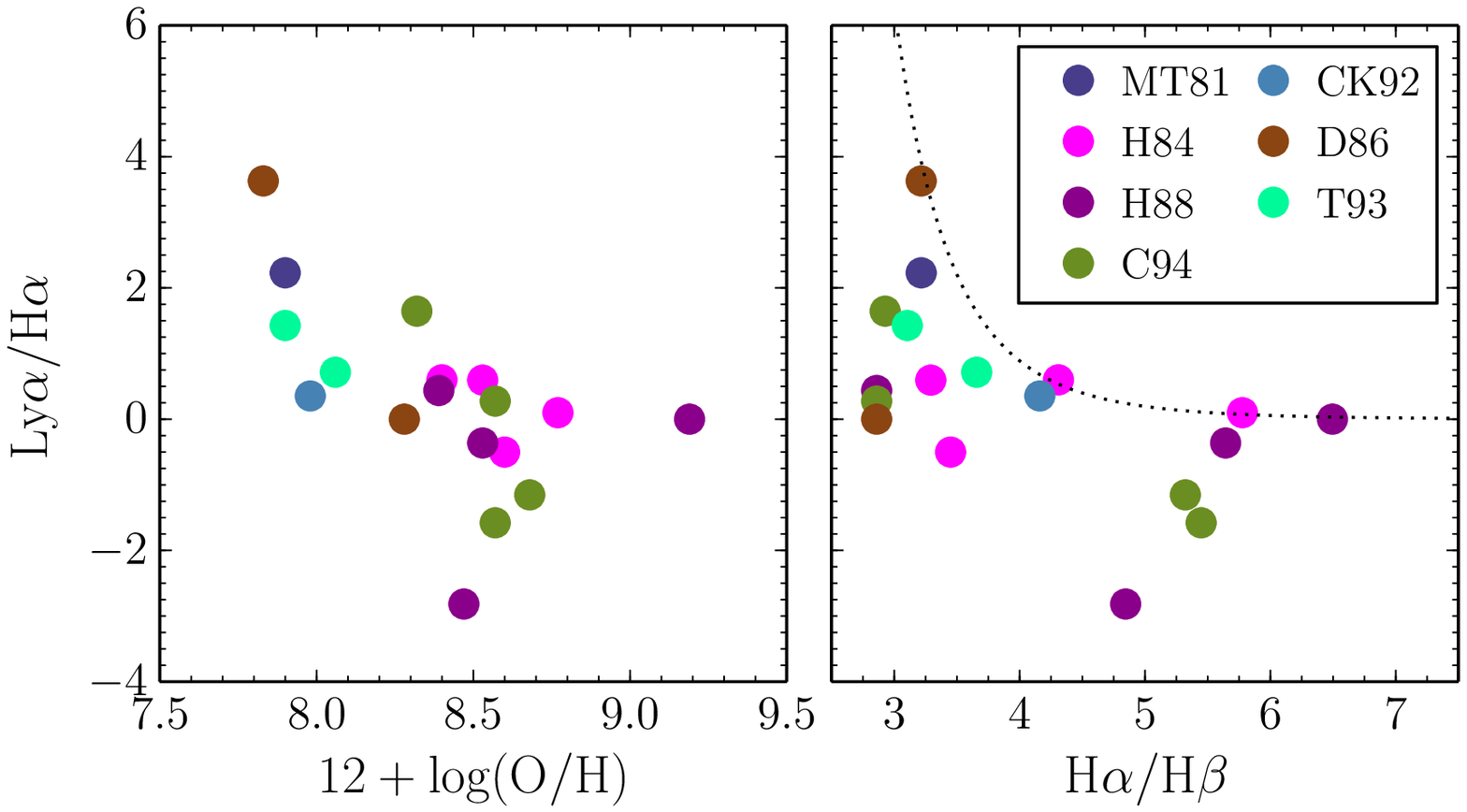}
\caption{The observed ratio of \lya/\halpha\ fluxes, shown on a linear scale
against nebular metallicity and the \halpha/\hbeta\ line ratio for the full
sample of star-forming galaxies observed with the IUE, and reprocessed by
\citet{Giavalisco1996}.  Negative \lya/\halpha\ ratios are the result of the
\lya\ feature being dominated by ISM absorption. The reference coding is
MT81=\citet{Meier1981}, H84=\citet{Hartmann1984}, H88=\citet{Hartmann1988},
C94=\citet{Calzetti1994}, CK92=\citet{Calzetti1992}, D86=\citet{Deharveng1986},
T93=\citet{Terlevich1993}.  The dotted black line shows the effect of the
\citet{Calzetti2000} dust attenuation curve, assuming intrinsic ratios of
\lya/\halpha=8.7 and \halpha/\hbeta=2.86 and no scattering.  Most of the
galaxies lie far below this curve.}\label{fig:gia96}
\end{center} 
\end{figure*}

\citet{Giavalisco1996} performed a full reanalysis of the IUE archival data,
resulting in several changes.  Firstly the IUE data reduction software reached a
higher level of maturity, and new spectral extraction and cosmic-ray removal
tools were implemented: this reduced the measured \ewlya\ significantly in about
half the sample, and completely removed the weak \lya\ feature that had been
reported in some galaxies.  Secondly, these authors performed proper spatial
matching between the IUE and apertures used for optical line spectroscopy.  When
the homogenized reanalysis was complete, both the \ewlya\ and \lya/Balmer line
ratios showed no significant correlation with nebular dust attenuation or the UV
continuum colour ($\beta$), and only a weak but significant correlation with
nebular oxygen abundance.  Results concerning the \lya/\halpha\ line ratios can
be seen Figure~\ref{fig:gia96}.  

Regarding these correlations, it is not clear why \lya\ throughput should be
more strongly correlated with oxygen abundance than nebular attenuation.
Metals alone do not absorb \lya, which can only happen by interaction with dust
grains.  Thus if the \ewlya--$Z$ anti-correlation results from a positive
correlation between metal and dust abundance, then a tighter correlation
between \lya/\halpha\ and \ebv\ would be expected.  This correlation is
completely absent.   Furthermore, when the dust reddening measured from the
Balmer decrement is used to correct the observed \lya\ for extinction, the
\lya\ flux does not reach the expected case B recombination value in a single
galaxy in the IUE sample.  This demonstrates that either some preferential
attenuation of \lya\ must be at play in every galaxy, or/and that dust
attenuation laws, when applied as a screen of absorbing material, are not
representative.  

When interpreting these analyses of the IUE samples, it is important to keep in
mind the selection functions by which the individual samples were established.
When effective telescope areas are small and exposure times need to be long, the
result is small samples.  Some of the blue compact dwarf (BCD) and \hII\ galaxy
studies were designed to study the analogues to `primeval' galaxies that are
undergoing their first phase of star-formation, and the samples did not include
galaxies with strong starbursting nuclei \citep{Hartmann1984}.  Yet the primeval
stages of galaxy formation, prior to the initial dust and metal production, are
expected to be short-lived because the first generations of supernovae will
enrich the local ISM on timescales of just a few Myr.  Consequently the bulk of
the galaxy population we can observe in the high redshift universe is likely to
be significantly more metal-enriched than that of primeval galaxies
\citep[e.g.][]{Pettini2002,Shapley2003}, unless observations catch galaxies in
very narrow time window.  Thus while providing very interesting astrophysical
laboratories, the samples are biased and not necessarily in a direction tuned to
the real importance of \lya: probing the faint population of normal galaxies at
$z>2$.  The IUE samples contain few galaxies that can be considered the local
analogues of high-$z$ Lyman Break Galaxies (LBG), \lya-emitters, or primeval
galaxy building blocks.

\section{HIGH RESOLUTION SPECTROSCOPY WITH THE HUBBLE SPACE TELESCOPE}\label{sect:hstspec}

\begin{figure*}[t!]
\begin{center}
\includegraphics[width=0.35\textwidth]{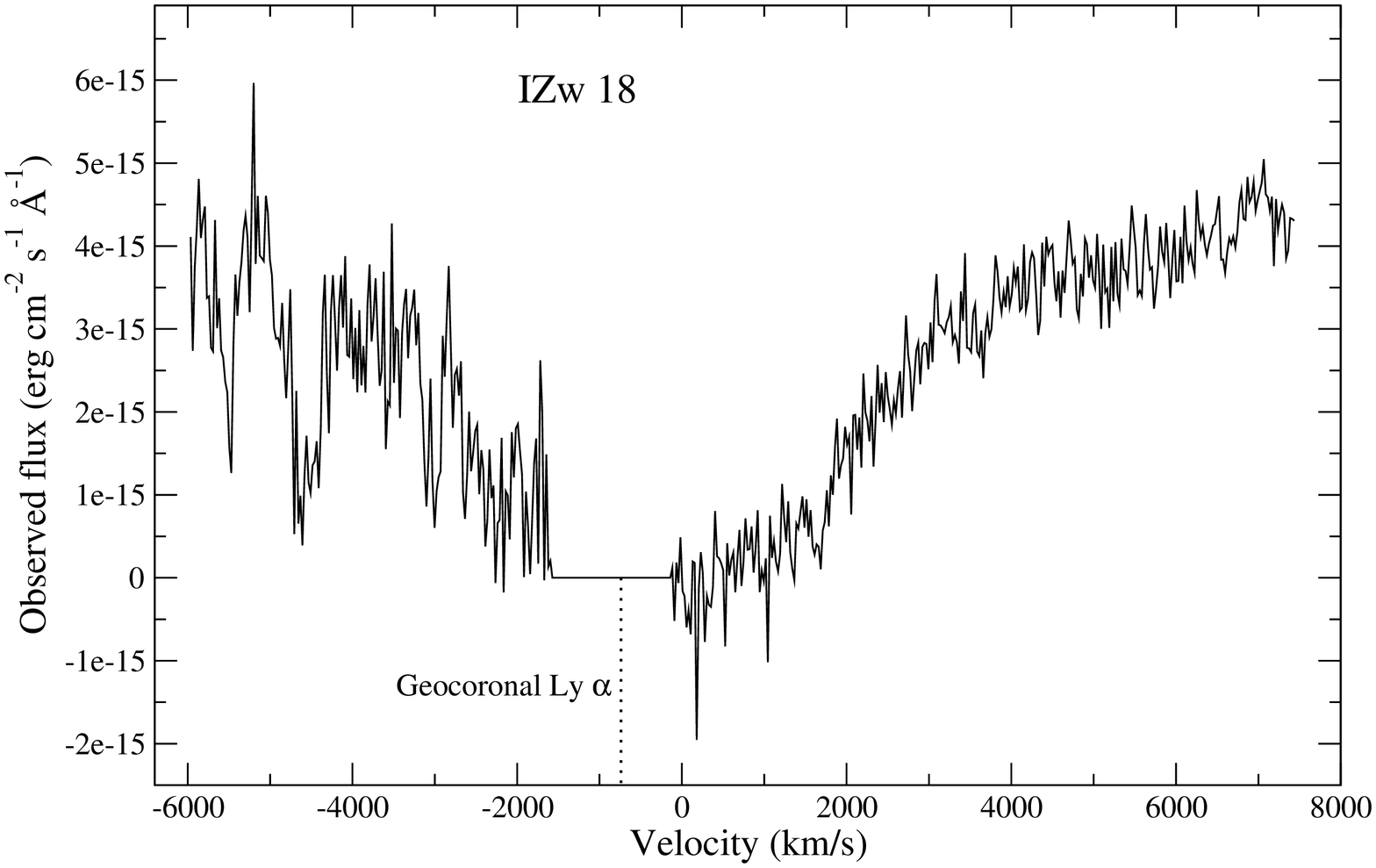}
\includegraphics[width=0.361\textwidth]{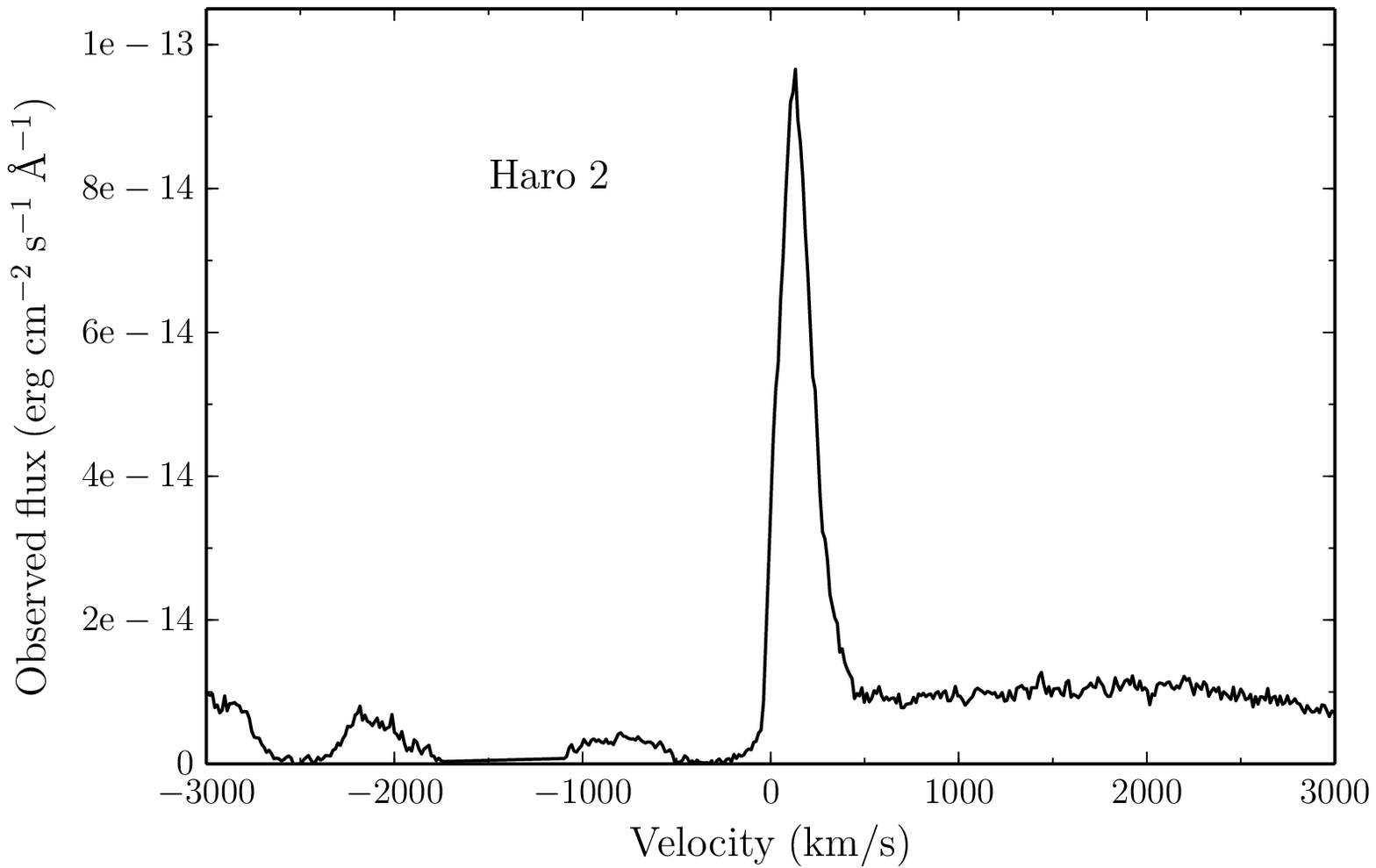}
\caption{The \lya\ spectra of first dwarf galaxies to be observed in the far UV
with HST.  The \emph{left} panel shows \izw, first observed with GHRS by
\citet{Kunth1994}. The geocoronal \lya\ line has been masked out and is shown by
a line set to zero flux at velocities of $-500$ to $-2000$~\kms.  No intrinsic
\lya\ emission is seen, and damping wings are visible that extend to at least
6000~\kms.  \emph{Figure is taken from} \citet{Mas-Hesse2003}.  The \emph{right}
panel shows Haro\,2, first observed by \citet{Lequeux1995}.  Haro\,2 has a much
higher dust abundance than \izw, but shows a strong \lya\ emission line.
Furthermore the line profile is \pcyg-like, which indicates the intrinsically
produced frequencies have been redistributed by scattering in an expanding
neutral medium.  \emph{Data are replotted from}
\citet{Mas-Hesse2003}.}\label{fig:dwarfs}
\end{center} 
\end{figure*}

High-resolution ultraviolet observations in the nearby universe, both imaging
and spectroscopy, are one of the major legacies of HST.  The main UV
spectrographs on HST were designed to have high resolving powers, which
necessitated either small entrance apertures or the use of narrow slits.  The
natural consequence is that while HST spectra may be rich with features, any
measurements strictly reflect the properties of the stars and gas that fall
within the aperture.  In most cases these will be the regions of highest UV
surface brightness, which in general will be unobscured massive stellar
clusters.  However it is not necessary that these local properties, whether
regarding \lya\ or interstellar gas, are representative of the entire galaxy.
Thus we must always keep in mind that HST derived measurements of gas
kinematics, densities, and covering, such as those discussed in this Section,
are local.

The first-generation instrument, the \emph{Goddard High Resolution Spectrograph}
(GHRS) increased the resolving power over that of the IUE ($R\sim 250$) by a
factor of 10--100.  This enabled studies of the kinematics and covering of the
atomic gas which are topics completely absent from discussion in every previous
paper presenting IUE data.  Furthermore, with the IUE observers needed to target
galaxies at high enough radial velocity to separate the intrinsic \lya\ feature
from the bright \lya\ line produced in the Earth's corona, which is brighter
than any known extragalactic object.  Indeed because of the tentative
anticorrelation between \ewlya\ and metallicity, \citet{Meier1981} already
commented upon how it was unfortunate that the lowest metallicity galaxy known
-- \izw, with $12+\log (\mathrm{O/H}) \approx 7.2$ -- had at too small a redshift to
be observed with IUE, for logically it must be very bright in \lya.  With
HST/GHRS it could finally be observed.

\subsection{HST Finds Deep \lya\ Absorption}\label{sect:bcd}

GHRS observations of \izw\ revealed a profile showing only damped \lya\
absorption and no hint of emission \citep{Kunth1994}.  The Lorentzian wings of
the absorption profile can be traced out to at least 6000~\kms\
(Figure~\ref{fig:dwarfs}; \citealt{Mas-Hesse2003}), implying a \hI\ column
density above several $10^{21}$~\percm.   Furthermore, measurements of the low
ionization stage (LIS) metal lines that form in the neutral ISM (e.g.
\oI~$\lambda1302$, \siII~$\lambda1304$) show that the \hI\ gas is static with
respect to to the \hII\ regions, and at the measured column density \lya\
radiation at line-centre will have to traverse $10^7$ optical depths in order to
escape if the \hI\ is homogeneous. 

Subsequent observations of \izw\ with the \emph{Space Telescope Imaging
Spectrograph} (STIS) enabled detailed, spatially resolved, and empirically
well-constrained studies of \lya\ radiative transport, which indeed shows that
this profile can be reproduced, including spatial variation in the damping
wings, using column densities of $N_\mathrm{HI} \sim 3\times 10^{21}$~\percm,
and \ebv=0.05 \citep{Atek2009izw18}.  This is fully consistent with the directly
observed values on both \nhi\ and \ebv.  Moreover the same transport simulations
predict that for static gas with this \hI\ column density, almost all the \lya\
radiation is absorbed by the small amount of available dust (escape fraction the
order of $10^{-4}$ to $10^{-3}$), as many scattering events increase the
probability of absorption. 

\begin{figure}[t!]
\begin{center}
\includegraphics[width=0.8\textwidth, angle=0]{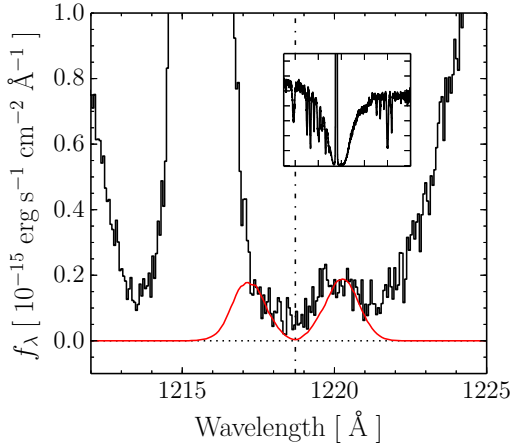}
\caption{Very deep \lya\ spectrum \izw\  obtained with the Cosmic Origins
Spectrograph.  While previously referred to as a \lya\ absorbing galaxy, \izw\
shows a very small but significant bump of \lya\ emission.  This bump is
redshifted from the expected systemic velocity (the dot-dashed vertical line)
derived from the \halpha\ line.  The overlaid red line shows the profile of
\lya\ transferring through a static shell of atomic hydrogen with a column
density of $\log(N_\mathrm{HI}/\mathrm{cm}^{-2})=21.1$, taken from the
\citet{Schaerer2011} grid of transfer models.  The strong emission feature is
the geocoronal \lya\ emission line.  The inset shows the wider spectral
profile.}\label{fig:izw18} 
\end{center} 
\end{figure}

With the HST servicing mission 4, the \emph{Cosmic Origins Spectrograph} (COS)
was installed on HST.  COS has a similarly sized entrance window to GHRS and
similar grating specifications, but can observe much larger wavelength range in
a single observation, and is many times more sensitive.  Recently obtained COS
observations of \izw\ enable us to go much deeper than was previously possible,
and a COS/G130M exposure of 29~ksec actually does reveal a small bump of \lya\
emission, hidden at the bottom of the absorption profile
(Figure~\ref{fig:izw18}; see also \citealt{Lebouteiller2013,James2014}).  The
\lya\ emission feature lies in the centre of the absorption trough, and the flux
amounts to just $\approx 4\times 10^{-16}$~\ergseccm.  Recalling at this point
that HST spectrographs measure only very local properties, we would need to make
a large aperture correction to estimate the total \lya\ output.  However
assuming an exponential \lya\ surface brightness profile we would need a scale
length of 80 arcsec in order for 100\% of the \lya\ to be emitted (integrating
to infinity).  While this is not ruled out -- \hI\ is extends over at least a
square arcmin \citep{vanZee1998} -- this scale length is 40 times the UV
effective radius, which is an extreme extension of \lya\ compared with other
nearby objects (Section~\ref{sect:hstimage}).  The alternative is that a large
fraction of the \lya\ photons are absorbed by dust after numerous scattering
events, as suggested by \citet{Atek2009izw18}.

In \izw\ the \lya\ bump is also offset from the systemic velocity (measured from
\halpha) by 350~\kms\ while the neutral gas shows bulk velocities of $-10$~\kms\
(measured from \siII).  The bulk motion is insufficient by an order of magnitude
to kinematically redshift the \lya\ by such a velocity, so other physical
processes must be at play.  The red line in Figure~\ref{fig:izw18} shows a
radiative transport model produced by the grid of \citet{Schaerer2011}, for a
completely static and dust-free \hI\ shell with a column density
$\log(N_\mathrm{HI}/\mathrm{cm}^{-2})= 21.1$, which is within 0.25 dex of
previous estimates based upon modeling just the absorption feature.  A plausible
interpretation is that the bump is the red half of a double-peaked profile,
which results from wing scattering events that shift photons in frequency many
Doppler widths into the redistribution profile.  If so, and the scattering
medium is completely static, there would be a corresponding blue peak at
--350~\kms; this however would be hidden below the geocoronal \lya\ line, that
swamps any intrinsic emission. 

Observations of similar dwarf galaxies show the deep \lya\ absorption seen in
\izw\ is not unique.  Two of the other most metal-deficient galaxies known,
\sbs\ and Tol\,65 \citep{Thuan1997}, have metallicities just a factor of 2
higher than \izw.  Both also show broad \lya\ absorption profiles with
equivalent widths of $-20$ to $-30$~\AA, no hints of \lya\ emission, and clear
damping wings that imply \hI\ column densities above $2\times 10^{21}$~\percm.
Similar deep absorption is visible in the COS spectrum of low-metallicity dwarf
SBS\,1415+437 \citep[][$12+\log(\mathrm{O/H})\approx 7.6$]{James2014}, which
also shows a similarly redshifted bump of \lya\ in emission.  These galaxies are
very rare in the local universe, and one may question whether far-reaching
conclusions may be drawn from them.  Nevertheless, such objects may become more
abundant at higher redshifts, and any complete theoretical picture of \lya\ must
also include them.

\subsection{Galaxy Winds and Atomic Gas Kinematics}

In contrast to the strongly absorbing dwarfs the second BCG with a published HST
\lya\ spectrum, Haro\,2, was found to emit a strong \lya\ line with
\ewlya$\approx 7$\AA\ \citep[emission part only,][\emph{right} panel of
Figure~\ref{fig:dwarfs}]{Lequeux1995}.  This is particularly curious because
Haro\,2 is an order of magnitude more dust- and metal-rich than the dwarf
galaxies.  Comparing with the absorbing BCDs discussed in
Section~\ref{sect:bcd}, the total column density of hydrogen along the
line-of-sight is roughly the same, but with two important differences. Firstly,
much less of that hydrogen column density is contributed by the neutral phase,
although the measured column of \nhi\ $\gtrsim 10^{20}$~\percm\ would still be
sufficient to cause a damped absorption.  Secondly, the absorption centroid of
\lya\ is blue-shifted by $\approx 200$~\kms\ relative to the systemic frame of
rest of the galaxy.  The result of this first resolved observation of a \lya\
emission line is an asymmetric profile that comprises a blue absorption
component and a red emission peak, similar to the \pcyg\ profile.  Even though
the wing of the absorption profile is quite extended, the velocity offset is
sufficient to shift the neutral medium partially out of resonance with \lya, and
enable some of the \lya\ radiation to escape.  

The LIS lines intrinsic to Haro\,2 were also found to be blue-shifted with
respect to the systemic velocity and by the same velocity as measured from the
\lya\ absorption ($\approx 200$~\kms).  The peak of the \lya\ emission, however,
is instead redshifted by 350~\kms, or roughly twice the blueshift of the neutral
medium.  This led \citet{Lequeux1995} to suggest that much of the \lya\ is able
to avoid \hI\ absorption in Haro\,2 because it does not see the atomic gas as
static, and the redshifted \lya\ emission supports a picture in which the \lya\
that is emitted is `backscattered' from a receding shell of \hI\ gas
\citep{Verhamme2008}.  Furthermore in Haro\,2, diffuse soft X-ray emission
covers and extends beyond the UV-bright, star-forming regions, which is produced
by the mechanical energy released by the star formation episode
\citep[][]{Oti-Floranes2012}.  This X-ray emission is spatially consistent with
an extension of the \lya\ emission in the 2D spectral image.  In contrast \izw,
which shows only \lya\ absorption and static low-ionization absorption lines, is
undetected at soft X-ray energies \citep{Ott2005}.  The conventional picture for
galaxy outflows is that cold gas is accelerated by expanding hotter gas
\citep[e.g.][]{Strickland2004}, which supports scenario where feedback-driven
outflows promote the emission of \lya.

\begin{figure*}[t!] 
\begin{center} 
\includegraphics[width=5cm, angle=0]{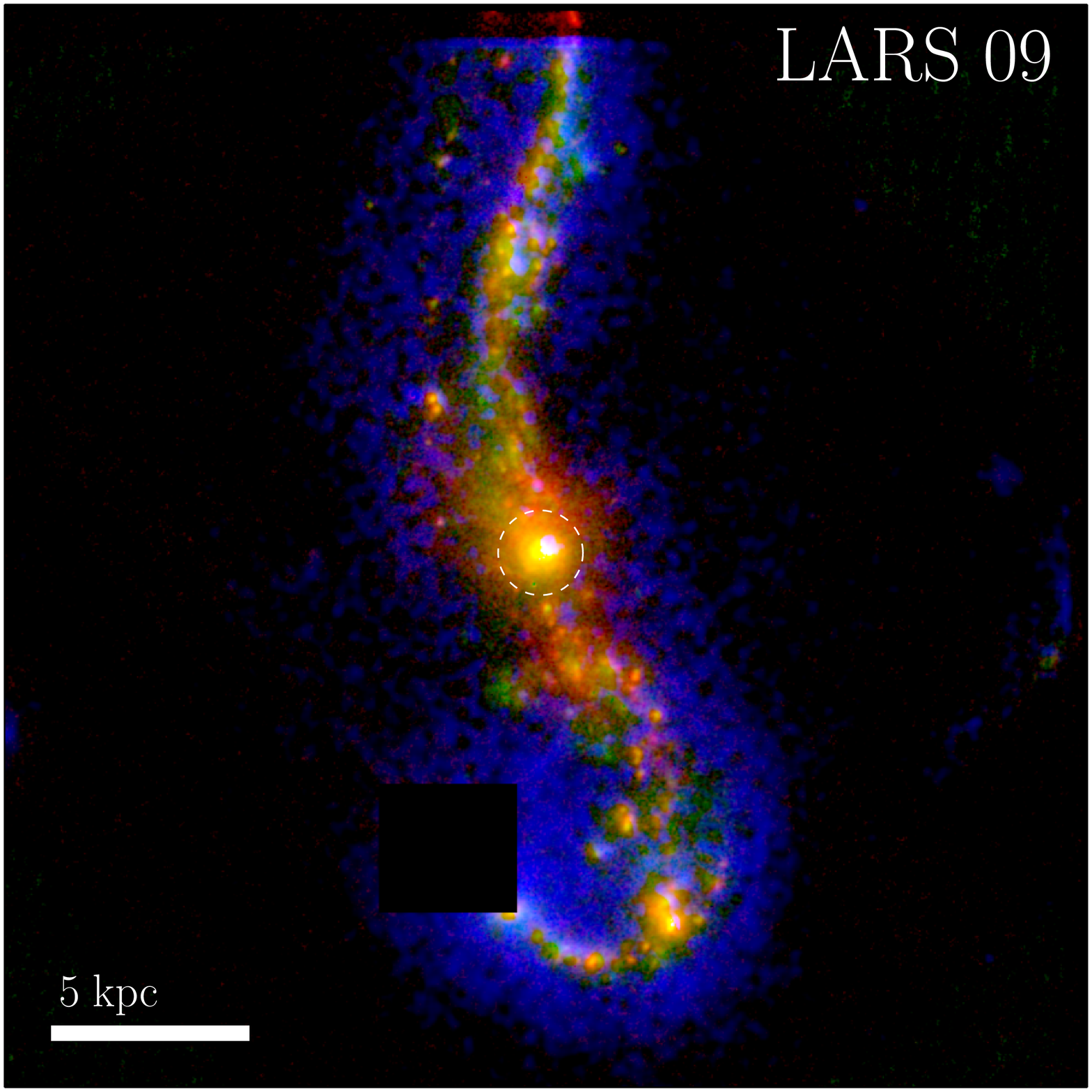}
\includegraphics[width=5cm, angle=0]{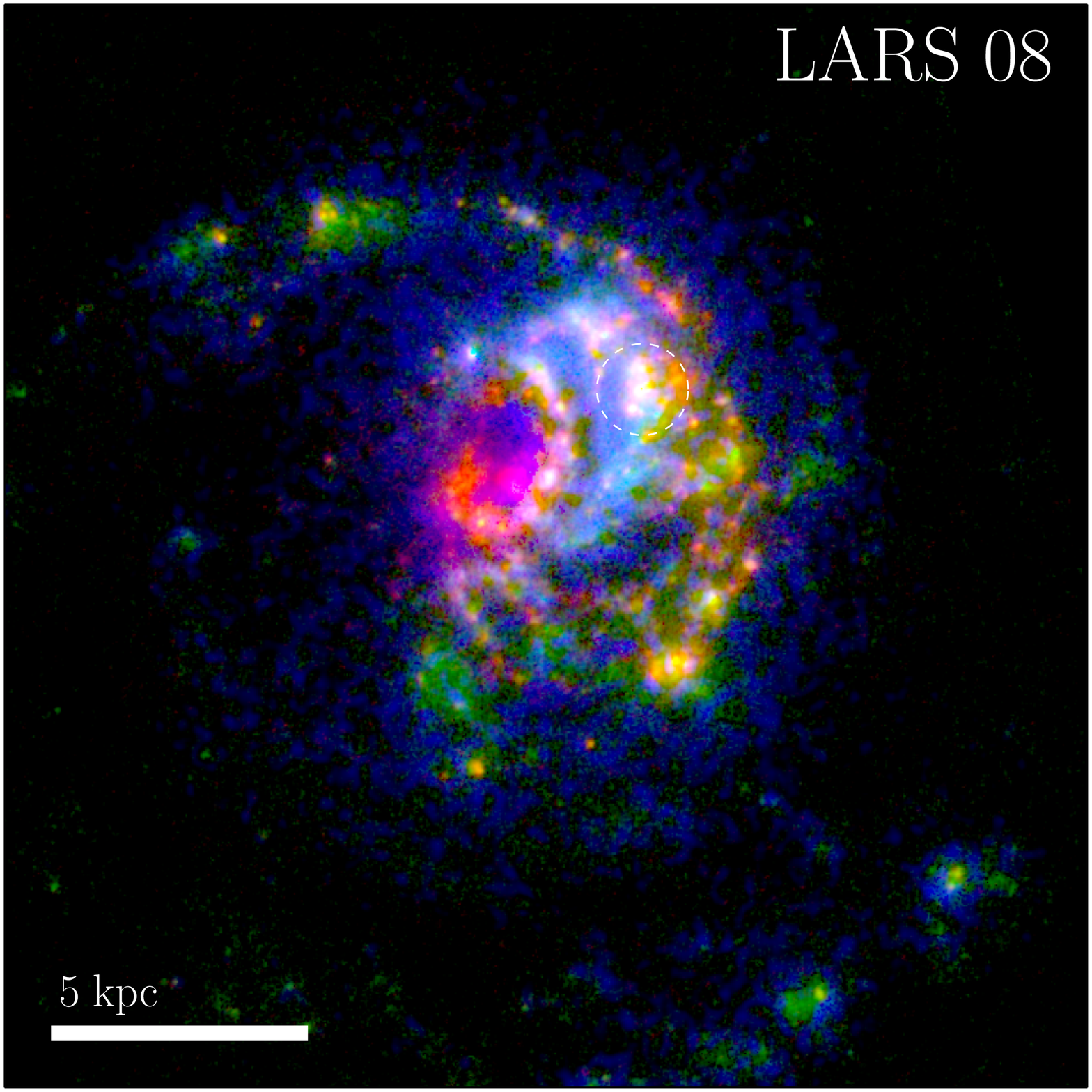}
\includegraphics[width=5cm, angle=0]{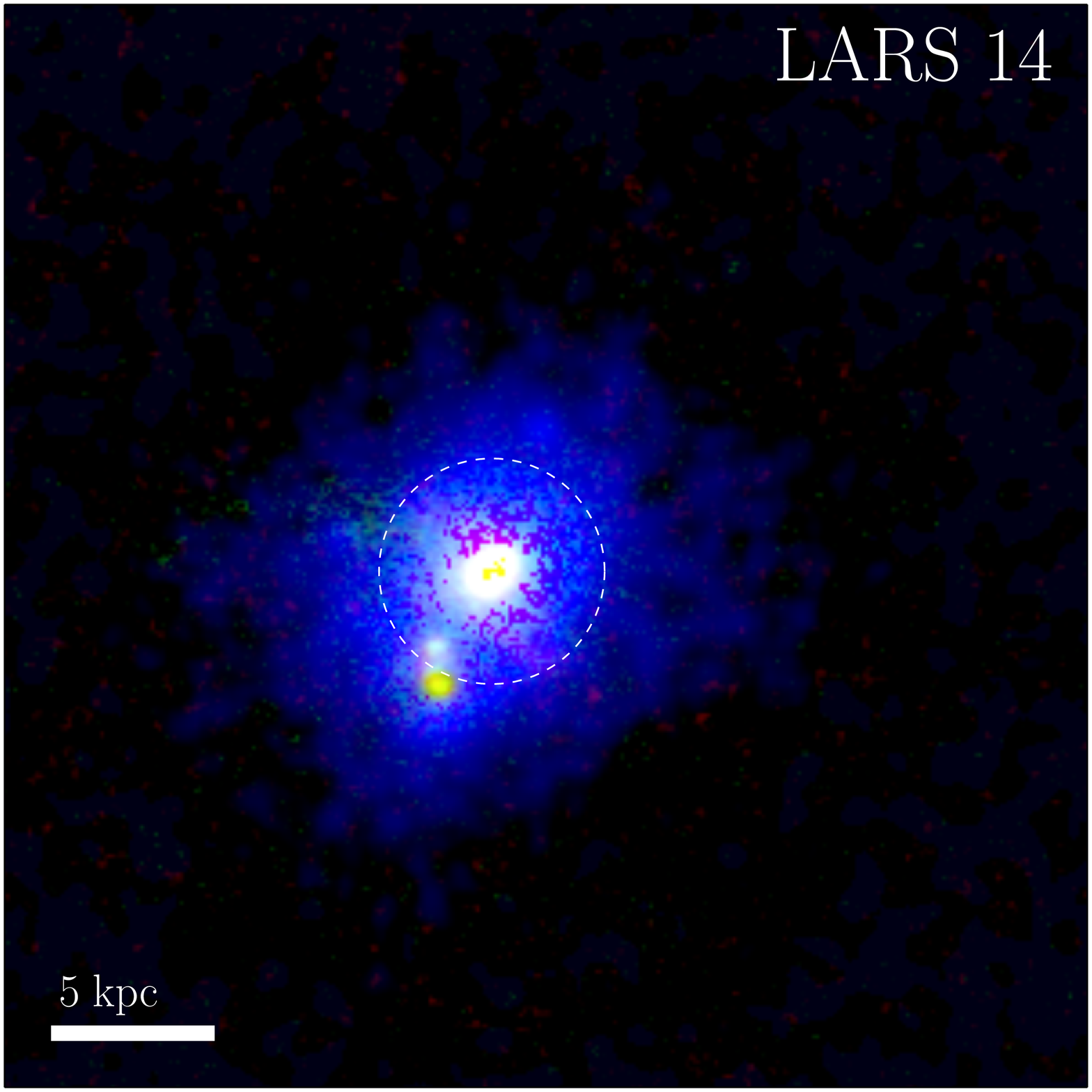}\\
\includegraphics[width=5cm, angle=0]{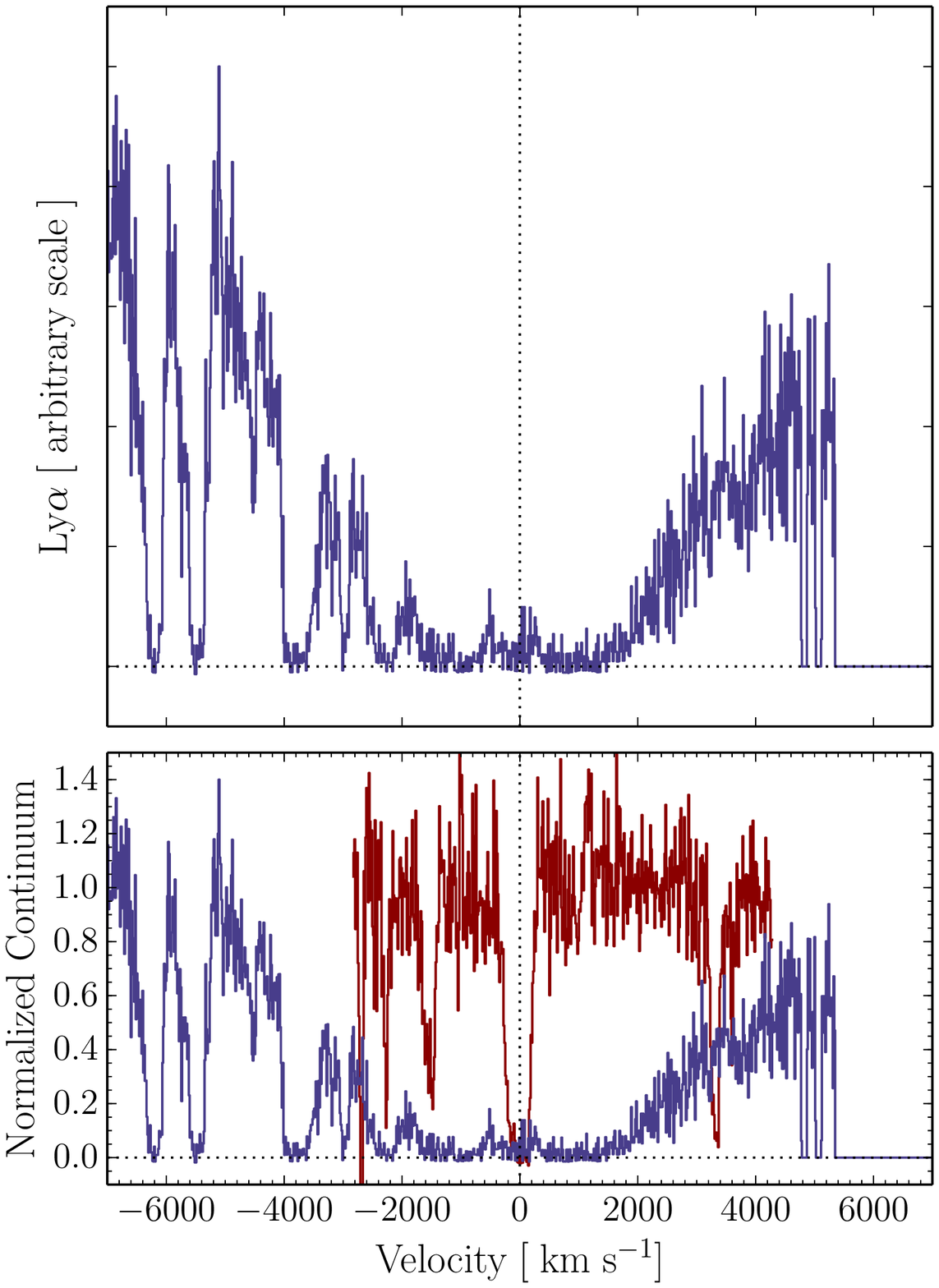}
\includegraphics[width=5cm, angle=0]{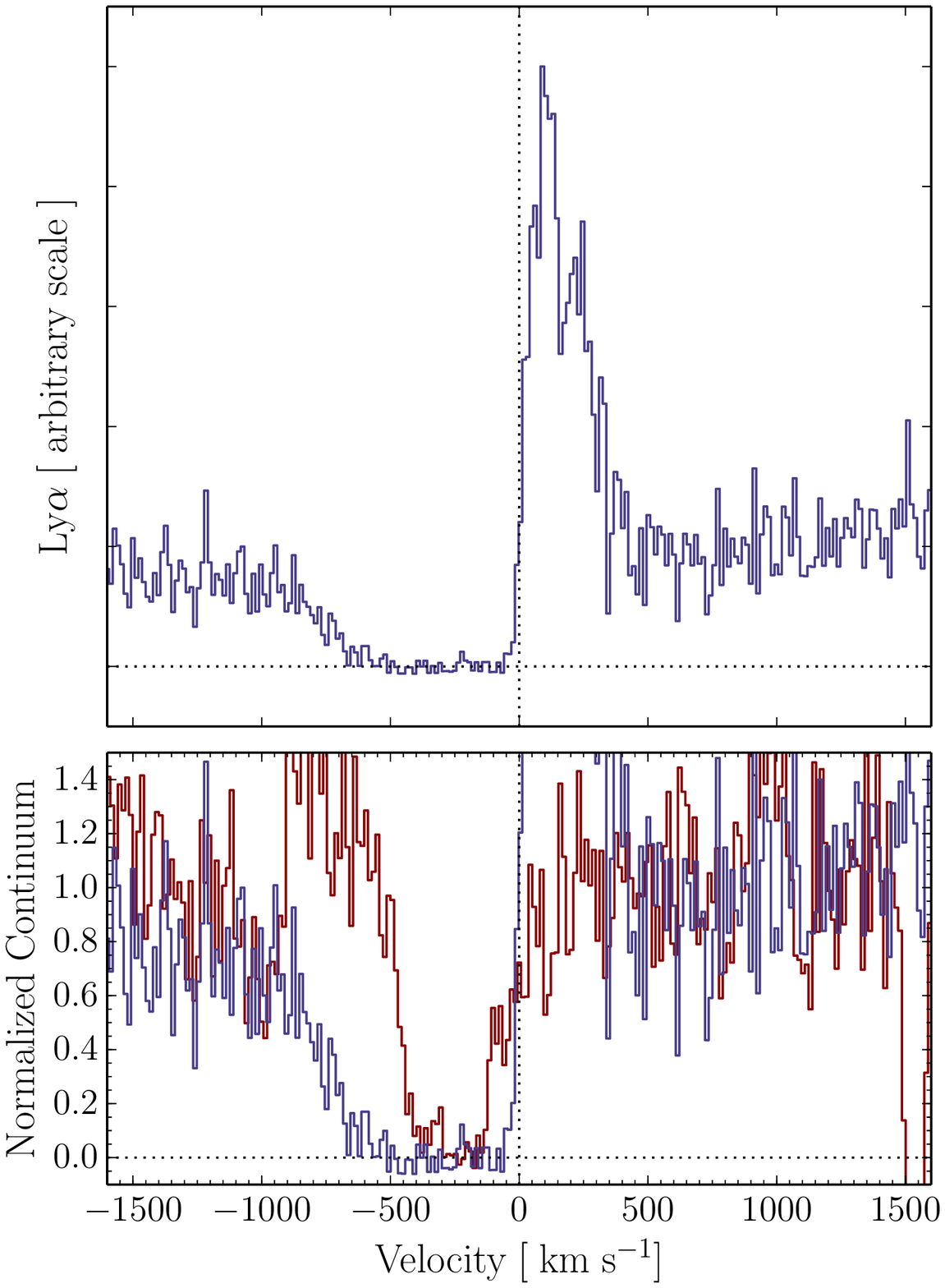}
\includegraphics[width=5cm, angle=0]{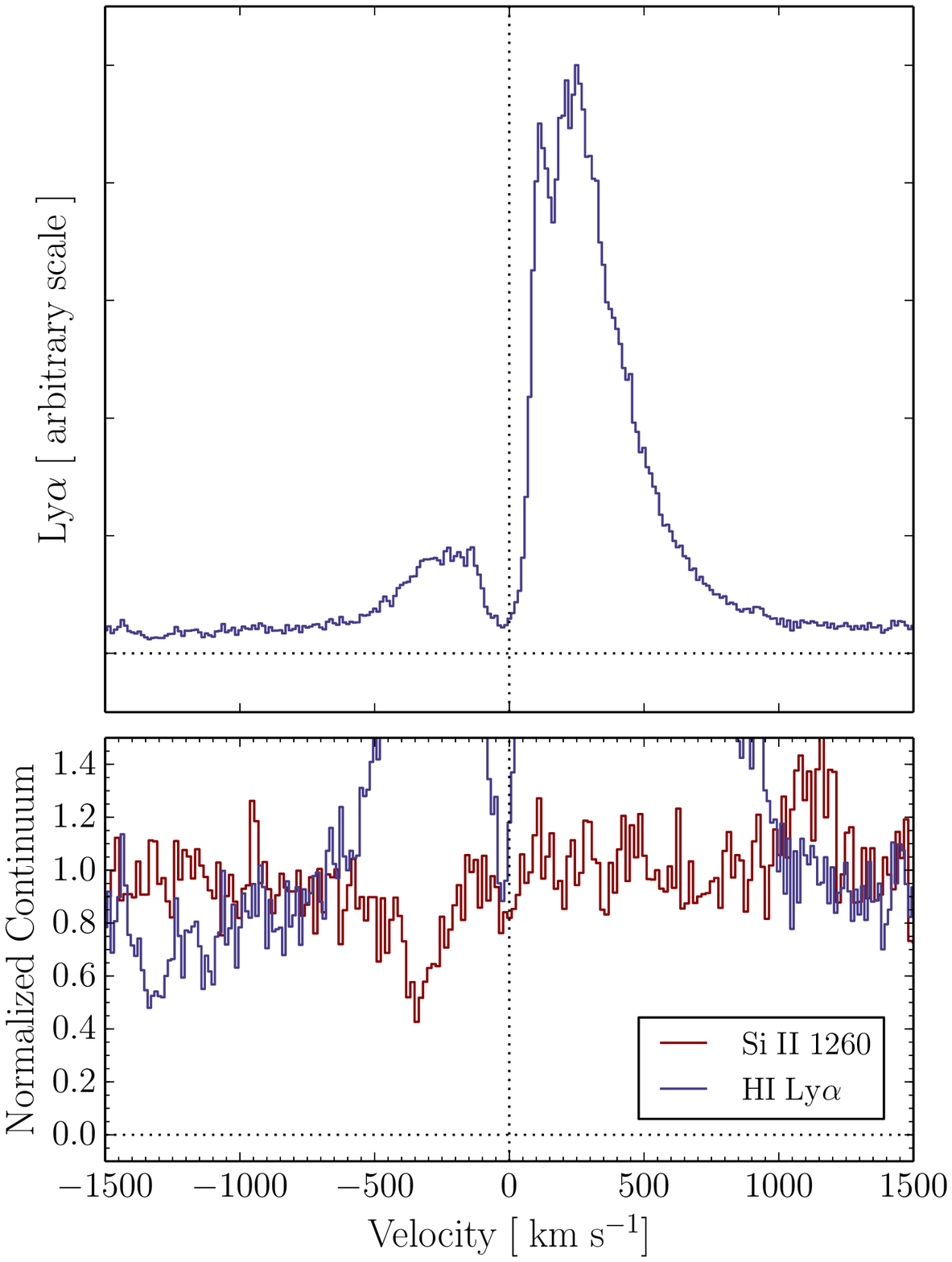}
\caption{Example images and \lya\ spectra of local galaxies.  \emph{Upper}
panels show colour composite images, encoding \halpha\ in red, FUV continuum in
green, and \lya\ in blue.  Dotted white lines indicate the size and position of
the COS aperture. \emph{Lower} panels show the corresponding spectra around
\lya\ (dark blue) and \siII$\lambda 1260$\AA\ (red).  Objects are selected to
illustrate the variety of \lya\ and LIS profiles.  Note the different velocity
scales, which extend to $\pm 7000$~\kms\ in the \emph{left}-most frame
but just $\pm 1500$~\kms\ in the other two.  The \emph{left} object, LARS\,09, shows a
completely absorbed \siII\ profile that is centred at zero velocity.  This
absorbing gas removes 300~\kms\ (FWHM) in the \siII\ line but completely damps
the \lya\ line, resulting in a FWHM that is 20 times broader.  The \emph{centre}
plot, LARS\,08, shows a similarly broad and saturated \siII\ profile, but one
that is offset in velocity by $\sim -250$~\kms.  The blue wing of the \lya\
absorption profile shows a similar shape and is offset in a similar way to the
\emph{left} panel, but in this case a redshifted \lya\ line is emitted.  The
\emph{right} panel shows a an example of a strong  \lya-emitter, in which the
\siII\ absorption is blueshifted by $\approx 350$~\kms, is far from saturated,
and a very bright \lya\ emission line is seen.}\label{fig:cosuvlya} 
\end{center} 
\end{figure*}

This early picture easily generalized in a sample of eight local BCGs observed
by \citet{Kunth1998}, four of which show net \lya\ in emission and four
absorption.  For those with net absorption, their \oI$\lambda 1302$ and
\siII$\lambda 1304$ absorption lines lie close to the systemic velocity (within
25~\kms), while the other four show outflowing gas with centroid velocities
shifted by 60--180~\kms\ \citep[see also][]{Leitherer2013}.  This correlation
does not necessarily imply a causal relationship and the fact that \lya\ is seen
to be locally emitted where outflows are strong could also be explained
also by orientation: transport models show more \lya\ to be emitted
perpendicular to galaxy disks purely because of the \hI\ distribution
\citep{Verhamme2012,Laursen2013}, and winds are also stronger in the polar
direction because the pressure is also lower
\citep[e.g.][]{Bland1988,Veilleux2002}. More compelling evidence for a causal
association comes from the fact that all the \lya\ emission lines in the sample
show \pcyg-like asymmetric profiles, indicating that photons are interacting
directly in the outflowing medium.  For this, it is much harder to argue for a
non-causal relation. 

Galaxies in the \citet{Kunth1998} sample were originally chosen to span a range
of metallicities and dust contents, but both of these quantities were found to
be secondary in governing \lya\ emission/absorption when compared to the
presence/absence of outflowing neutral gas.  

COS \lya\ observations of local galaxies are ongoing, but already the instrument
has far outdone the GHRS in terms of numbers.  Using larger samples of both
FUV-selected \citep{Heckman2011} and \halpha-selected objects
\citep{Wofford2013} this picture of kinematic regulation easily has
strengthened.  For galaxies in the `Lyman-break analog' (LBA) samples of
\cite{Heckman2011} \pcyg\ emission is ubiquitous while for the \halpha-galaxies,
redshifted \lya\ peaks and LIS lines blueshifted by around 100~\kms\ are found
among the emitters, while absorption lines (including \lya) consistent with zero
velocity shift at 68\% confidence are exhibited by the absorbers.  

Similar results are found among the \emph{Lyman alpha Reference Sample} (LARS;
Section~\ref{sect:halos}; \citealt{Rivera-Thorsen2015}), which are summarized in
the \emph{lower} panels of Figure~\ref{fig:cosuvlya}.  The \emph{left}-most
panel shows an example where the ISM in front of the brightest nuclear star
cluster (where the COS aperture is positioned) is static, and from where broad
damped \lya\ absorption is also observed.  From other regions of the galaxy,
however, \lya\ emission is recovered and the galaxy becomes a weak \lya\ emitter
(\ewlya~$\approx 10$\AA) in apertures that encompass the galaxy.  The
\emph{central} panel shows an example where the neutral ISM is outflowing along
the line-of-sight by around 250~\kms, the \hI\ absorption is similarly
blueshifted, and a weak \lya\ emission feature is visible within the pointing of
the COS.  The \emph{right} panel instead shows a galaxy where the atomic gas is
outflowing at higher velocity still, and a very bright \lya\ emission line is
visible with \ewlya~$\approx 40$\AA\ (80~\AA\ when including extended emission).

\begin{figure}[t!] 
\begin{center}
\includegraphics[width=0.9\textwidth,angle=0]{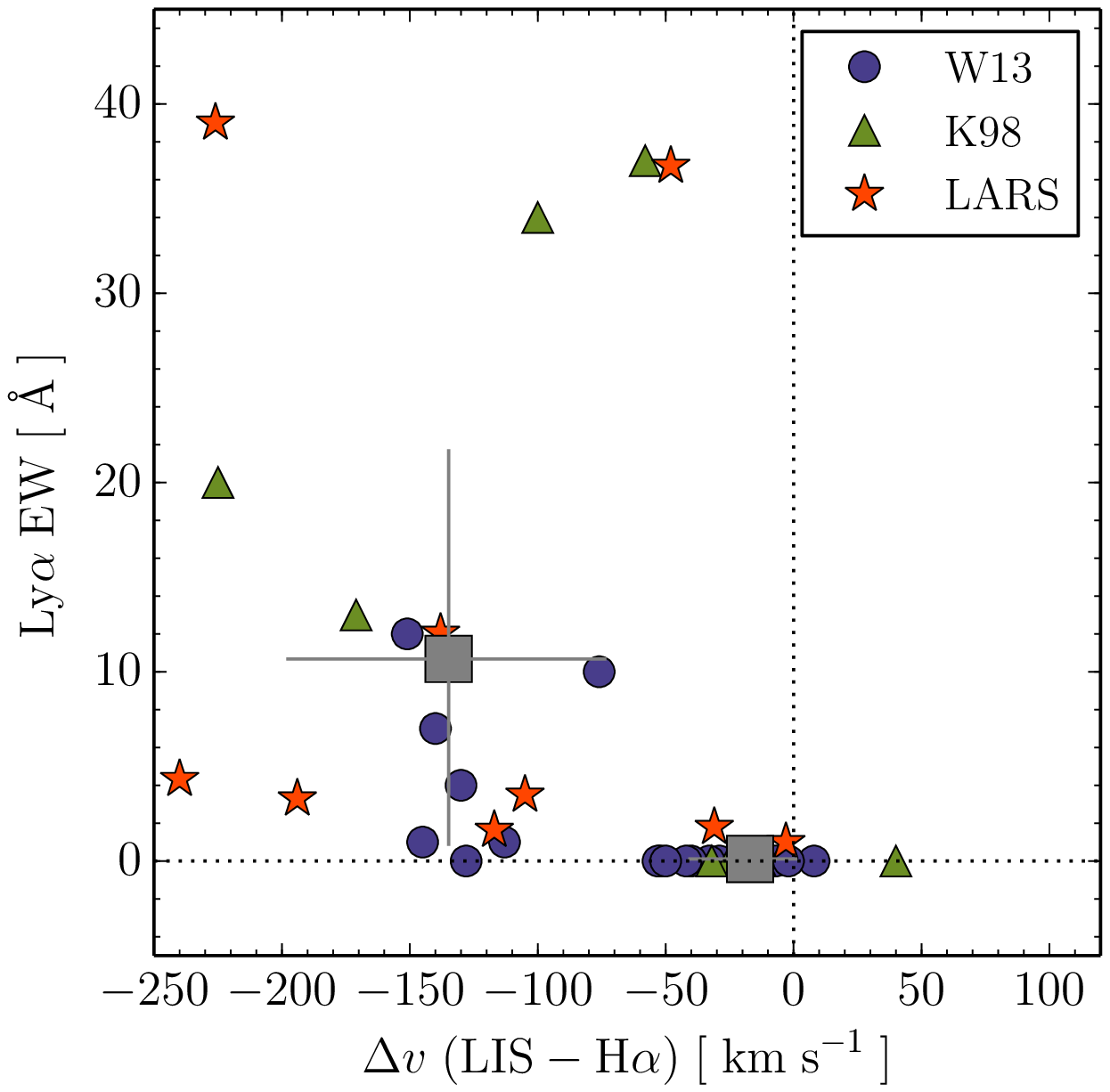}
\caption{The relationship between \ewlya\ and outflow velocity of the atomic
gas.  All measurements are made within spectroscopic apertures with HST/GHRS and
COS.  Where the \lya\ absorption component dominates over emission \ewlya\ is
set to zero.  $\Delta v$ is calculated from the average of velocity centroids of
all observed LIS lines compared with the intrinsic velocity of the nebulae
measured from optical emission lines.  A cut of $\Delta v=-50$~\kms\ divides the
sample in half by $\Delta v$, segregating galaxies with fast outflows from
those with weakly outflowing, static, or inflowing neutral media.  Grey points
with errorbars represent the average and standard deviation of galaxies of these
two sub-samples.}\label{fig:ewdv} 
\end{center} 
\end{figure}

Obviously if all neutral gas can be cleared from zero velocity, then \lya\
should escape unhindered.  For the column densities probed using the \siII\
lines, this appears to be the case in the right panel of
Figure~\ref{fig:cosuvlya}, where only more tenuous gas appears to remain at
$\Delta v\approx 0$ to absorb \lya.  Yet removal of cold gas from $\Delta
v\approx 0$ not a requirement for \lya\ emission, and \citet{Rivera-Thorsen2015}
present several examples of galaxies that show strong \siII\ absorption at
$\Delta v=0$, but also significant \lya\ emission (their Figure~8).  However
while there is clearly gas that does not have a velocity shift, the centroids of
the absorption profiles are offset -- usually by $\lesssim -50$~\kms\ -- which
demonstrates that there is fast-moving gas that can Doppler shift \lya\ out of
resonance with the static material.  Additional support for this comes from the
fact that the peak of the \lya\ profile is redshifted in every case.

Figure~\ref{fig:ewdv} summarizes the situation, by showing the average velocity
shift of low-ionization absorption lines compared with \ewlya\ measured in the
same aperture (diameter of 1.9 and 2.5 arcsec with GHRS and COS, respectively).
Outflow velocities span a range from $-250$ to $+50$~\kms\ (net inflow), with a
median value of $-50$~\kms.  No galaxy with low-ionization gas moving in the
velocity range $-50$ to $+50$~\kms\ shows significant \lya\ emission: all but
two of these objects show either net absorption or emit \lya\ with a total EW
below 3~\AA.  However galaxies with faster outflowing gas show a wide range of
\ewlya, which reaches up to 40\AA, with an average near 10\AA. We can state with
confidence that at least on small scales \lya\ emission is correlated with
feedback of sufficient magnitude, that acts to accelerate the highest density
neutral ISM along the line-of-sight.  Note, however, that this does not mean it
is the case that all galaxies with a strong outflow are \lya\ emitters --
clearly there are galaxies with $\Delta v\approx -150$~\kms\ and \ewlya\ below
2~\AA.  Furthermore while the correlation exists, the causal mechanism by which
feedback affects the transport has not necessarily been established.  For
example, whether feedback is simply shifting the \lya\ out of resonance by
scattering from bulk-flowing gas, or the instigation of fluid instabilities that
disrupt the ISM.

\subsection{Dissecting the Neutral Medium in Detail}

In addition to gas kinematics, the plethora of resonance absorption lines in the
UV also provide a proxy for the fractional covering of the cold \hI\ medium
\citep[e.g.][]{Savage1996,Pettini2002}.  The principle is simple: strong
resonance transitions are assumed to be saturated at normal metallicities and
column densities, and thus if the absorbing line does not drop to the level of
zero intensity then the observation hints that there may be multiple clouds
inside the spectroscopic aperture that do not fully cover the stellar sources of
continuum radiation that lie beneath.  Thus if there are direct sightlines
between the observer and the ionized regions, \lya\ may escape unimpeded, and
importantly, without frequency shift.  Indeed if scattering could be completely
mitigated and dust confined to the cold gas phase, \fesclya\ should be at least
$1-f_\mathrm{c}$, where \fc\ is the \hI\ covering fraction.

These methods have been used to place indirect limits on the escape of ionizing
radiation from starburst galaxies \citep{Grimes2009,Heckman2011}, and have
recently been verified by direct observations in the ionizing continuum with
HST/COS \citep{Borthakur2014}.  Similar tests, verified against \lya\ emission,
have been conducted at high-$z$ \citep[e.g.][]{Jones2013} and low-$z$ COS
studies focussed on \lya\ recently been presented.  For the majority of
local UV-selected galaxies the depth of the normally-saturated \siII\ lines
indicates a covering fraction close to unity.  However there is some deviation
from this: a weak trend is seen for galaxies with \fesclya\ above 0.1 to be
drawn from systems with \siII$\lambda 1260$~\AA\ absorption lines that are not
saturated \citep{Rivera-Thorsen2015}.  Very well exposed continuum observations
are needed to solve for covering fraction, but solutions include the possibility
of $f_\mathrm{c}<1$ for the thickest gas neutral columns in \lya-emitting
galaxies but not in the case of absorbing systems. 

Many of the resonant UV transitions, including the \siII\ discussed above, have
an associated fluorescent transition at longer wavelength, denoted with a *
(e.g.  \cII*), that provide additional diagnostics  of the atomic medium
\citep{Prochaska2011,Rubin2011,Jaskot2014,Scarlata2015}.  Since the absorption
lines are resonant, they may be partially filled by scattered radiation, very
similarly to \lya.  However, the fluorescent transition associated with each
line has a roughly similar Einstein $A$ coefficient to the resonant
de-excitation, which implies that roughly half of absorbed photons should be
emitted in the longer wavelength * line at every scattering.  In a symmetric,
energy-conserving system without losses, absorption along the sightline must be
balanced by isotropic fluorescent emission.  

\citet{Jaskot2014} recently presented spectra of two particularly interesting
bright \lya\ emitters, with \ewlya\ between 70 and 150 \AA.  Both of these
objects are well-detected in the stellar continuum, but absorption lines of
\cII$\lambda 1334$ and \siII$\lambda 1260$\AA\ lines are barely visible.
However the fluorescent counterpart of each transition is clearly seen in
emission, suggesting that the ISM is indeed partly covered or shows a low \hI\
column density in these galaxies.  A similar spectrum is that of LARS\,14,
illustrated in the \emph{lower-right} panel of \ref{fig:cosuvlya}
\citep{Rivera-Thorsen2015}, which shows a bright \lya\ line with a blue peak,
incomplete \siII\ absorption, and a fluorescent emission line (seen at relative
velocity of +1200~\kms). 

Further information may be inferred from the profiles of \lya.  While \lya\ and
\siII$\lambda 1260$ have cross sections of the same order of magnitude and
become optically think at similar column density, the metal abundances imply
that \lya\ may be absorbed by gas that is not visible to metal absorption.
Particularly in the galaxies of \citet{Jaskot2014} and LARS\,14, the \lya\ lines
do not resemble the strongly asymmetric absorption+emission of \pcyg\ profiles
that occur at high column densities of completely covered gas.  Instead they
show double-peaked profiles with narrow absorption at $\Delta v=0$.  This
indicates that there must be absorbing \hI\ that is not Doppler shifted, but
that is also not of sufficient column density and/or metallicity to be seen in
metal absorption lines.  This implies \nhi\ between $10^{15}$ and
$10^{18}$~\percm\ for normal ranges of metal abundance.

\section{FROM SMALL TO LARGE SCALES: COMPLETING THE QUANTITATIVE PICTURE WITH
HST IMAGING}\label{sect:hstimage}

Are HST spectroscopic studies are performed in small apertures or narrow slits.
Even for the most distant objects discussed so far, the 2.5 arcsec entrance
window of COS corresponds to a physical size of just 6~kpc, and obviously the
apertures will sample much smaller scales in more nearby galaxies (e.g. just
130~pc in \izw).  Thus while providing a very rich picture about \lya\ and the
ISM, spectroscopic measurements are restricted to the chosen sightlines: as
shown in Figure~\ref{fig:cosuvlya}, these will sample only a fraction of the
galaxies.  These small apertures are necessary to get the high spectral
resolution, but to capture a representative fraction of the \lya\ they are
likely far too small, particularly when considering that \lya\ can scatter.  To
get this, spectroscopic observations must be complimented with large-aperture
imaging.   

Thick columns of neutral gas have been observed in most of the starbursts
mentioned so far, and spectroscopic data strongly suggest that scattering
removes \lya\ from the line-of-sight.  An important question becomes whether the
bulk of the absorbed \lya\ is truly absorbed, or simply scattered to larger
radii from where it is subsequently emitted.  Such information is vital for the
comparison with high-$z$ data, where ordinarily a flux and EW measurement may be
available but little more, and standard practice is to adopt small aperture
appropriate for point sources ($\sim 1-2$~\arcsec).  

\begin{figure*}[t!]
\begin{center}
\includegraphics[width=0.75\textwidth]{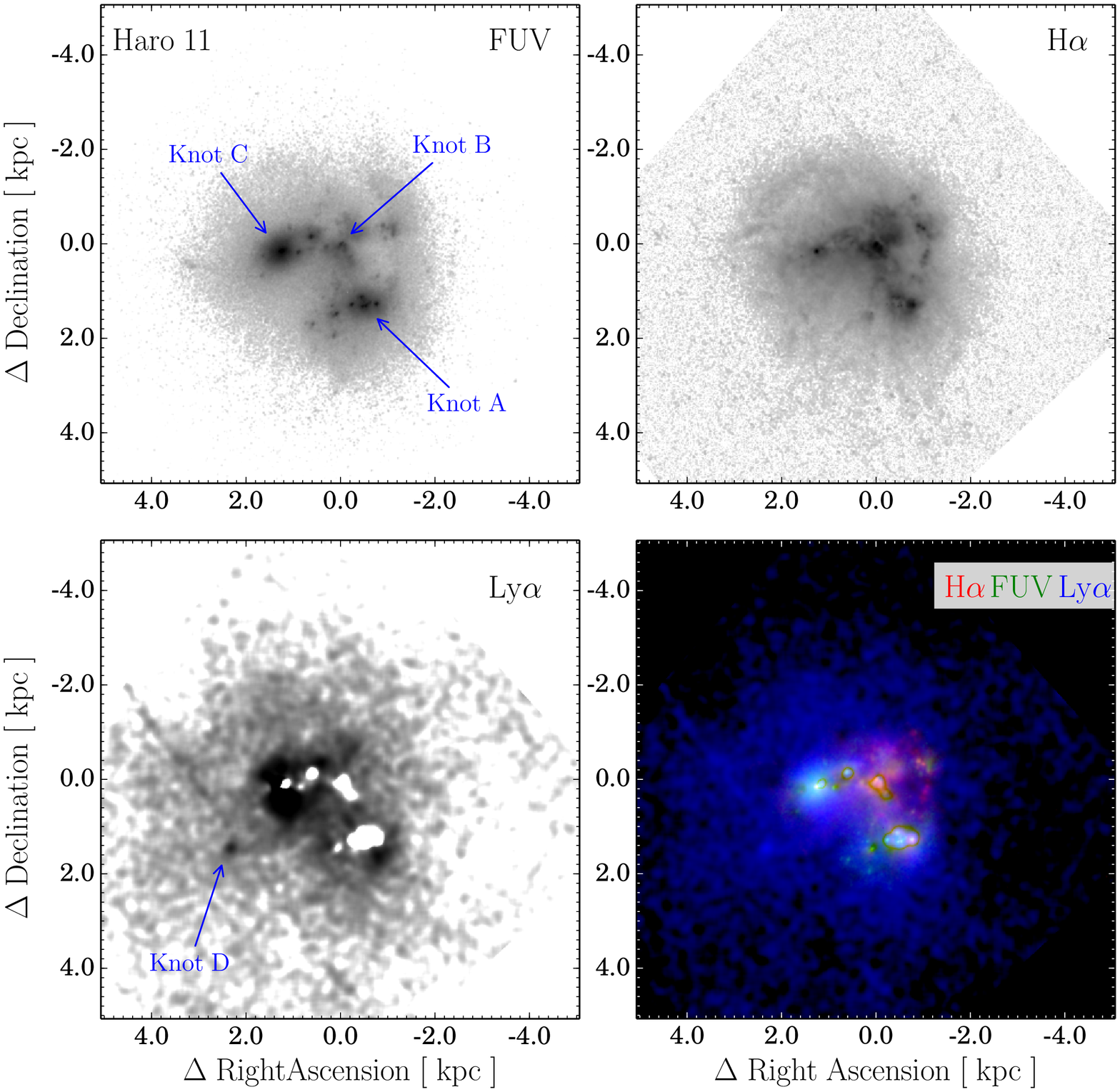}
\caption{HST/ACS imaging of luminous blue compact galaxy \har\
\citep{Kunth2003,Hayes2007}. \emph{Upper Left}: far UV continuum; \emph{Upper
Right}: continuum-subtracted \halpha; \emph{Lower Left}: continuum-subtracted
\lya.  The \emph{Lower Right} panel shows a composite of the other three images,
with \halpha\ in red, FUV in green, and \lya\ in blue.  The physical scale in
kpc is shown on the axes.  In the FUV image, the three main star-forming
condensations (knots A, B, and C) are labeled in the nomenclature of
\citet{Vader1993}; the \lya-emitting clump with no detection at any other
wavelength \citep[Knot D in][]{Kunth2003} is labeled in the \lya\
image.}\label{fig:h11}
\end{center} 
\end{figure*}

\lya\ imaging at $z=0$ became efficient with HST when the \emph{Advanced Camera
for Surveys} (ACS) was installed.  Although it was technically possible with
earlier cameras, the \emph{Solar Blind Channel} of ACS has a total throughput
of a few percent, and made such observations feasible for the first time.  Not
only did this bring about the first resolved information on \lya\ (aside from
very extended high-$z$ \lya-blobs), but also an immediate leap to an angular
resolution of $\approx 0.1$~arcsec.  However galaxies still need to be
sufficiently redshifted in order for \lya\ not to be absorbed my Milky Way \hI\
($cz \gtrsim 2500$~\kms\ is sufficient) so obvious targets such as very
well-studied nearby systems such as M82 or M33 still cannot be observed.

The first \lya\ images revealed a range of morphologies.  Near the central
starbursts, \lya\ is seen in both emission and absorption, where it may vary
between the two on sub-kpc scales with little obvious dependence upon local
properties such as age or reddening.  For example the first starburst with a
photometrically calibrated \lya\ image, \eso\ \citep{Hayes2005}, shows a lane of
\lya\ absorption that runs approximately E--W, and loops around one side of the
galaxy only.  This is seen at no other wavelength.  In other regions jets of
brighter \lya\ fuzz are visible, again seeming uncorrelated with obvious signs
of \lya\ production such as \halpha\ emission.  Obviously spectroscopic results
will be a strong function of aperture placement.  Furthermore, as soon as ACS
was turned towards local starbursts, \lya\ halos were discovered to surround the
starbursting regions \citep{Atek2008,Ostlin2009}.  I now proceed to discuss the
results of small-scale resolved analyses and extended \lya\ halos, beginning
with a case study of one system.

\subsection{Resolved Analyses: A Case Study of Haro\,11 }

Local luminous BCG \har\ is a Lyman break analog \citep[e.g.][]{Grimes2007}, and
emits \lya\ with a total EW of 15~\AA.  As shown in Figure~\ref{fig:h11}, it
comprises three main star-forming knots (labeled in the FUV image), all of which
are bright in the UV and \halpha, but only one of these condensations 
emits \lya\ \citep{Hayes2007}.  Following this breakdown of the galaxy, knots A
and B (the west-most two) are by far the brightest in \halpha\ and must
produce the bulk of the \lya\ radiation, but both absorb at \lya.  In contrast
it is only the single easterly knot C, which is the faintest of the three in
\halpha\ but brightest in the UV, that locally emits its \lya.  \har\ also emits
a halo of \lya\ emission, centred around knot C, which can most easily be
explained if \lya\ is re-radiated after scattering in the surrounding neutral
gas.  Diffuse emission also surrounds the two easterly knots, but at the
positions of the clusters absorption outweighs the emission, giving a negative
overall flux.  While the \lya\ surface brightness of the halo is low, it is also
very much larger than the UV continuum-bright regions, and in total contributes
$\approx 90$~\% of the total \lya\ flux \citep{Hayes2007}.  Results inferred
from small-aperture spectroscopic observations will be a strong function of both
the size and placement of the aperture. 

In \har, \lya\ produced in knots A and B may still be emitted, and all the
observation can say is that more radiation from the stellar continuum is
absorbed locally than the sum of directly emitted \lya\ and any \lya\ scattered
into the line-of-sight.  Thus for a given pixel we still cannot say whether
\lya\ is scattered and absorbed by dust locally, or whether it propagates some
kpc and contributes to the halo emission.  Remarkably in this three-region
decomposition of \har, is that the strongly emitting knot C is also the
dustiest, and shows $E_{B-V}\approx 0.4$ magnitudes, while knot A in particular
is far less extinguished \citep{Atek2008}.   Under the simplistic assumption
that dust plays a dominant role in regulating \lya\ visibility, this would be
unexpected, although results from these \lya\ absorbing knots are reminiscent of
the dwarf galaxies discussed in Section~\ref{sect:hstspec}. 

Similar phenomena were noted throughout the sample of nearby galaxies first
observed with ACS.  Specifically \eso\ was also found to exhibit a diffuse
\lya\ halo that dominates the \lya\ output \citep{Hayes2005}, and in a small
sample of six local starbursts, global \lya\ emission is invariably
associated with large-scale halo emission \citep{Atek2008,Ostlin2009}.

\begin{figure}[t!]
\begin{center}
\includegraphics[width=0.75\textwidth, angle=0, clip=true, trim=0mm 0mm 0mm 0mm]{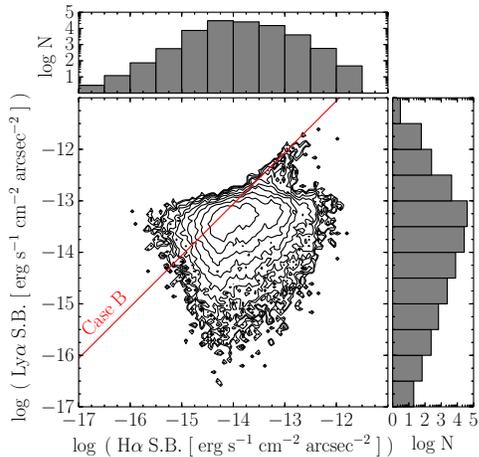}
\caption{Comparison of \lya\ and \halpha\ surface photometry for \har.  The
abscissa shows the logarithmic \halpha\ surface brightness, measured in
individual pixels in the HST images, while the ordinate axis shows the
corresponding surface brightness in \lya.  I.e. contours show the density of
points when comparing pixel values in the frames from Figure~\ref{fig:h11}.  The
red line shows the case B recombination line ratio of \lya$=8.7\times$\halpha;
intrinsically astrophysical nebulae produce \lya\ and \halpha\ radiation that
follow this line.  The plot is logarithmic, so only \lya\ emitting regions can
be visualized.  Histograms above and to the right show the overall the
distribution of light emitted as a function of surface
brightness.}\label{fig:px2px}
\end{center} 
\end{figure}

Nebulae produce \lya\ radiation intrinsically at an intensity of 8.7 times that
of \halpha\ (Section~\ref{sect:strength}), but when we contrast the local
surface brightnesses of the two lines a very wide range of line ratios is found.
These are illustrated in Figure~\ref{fig:px2px}, which contrasts the \halpha\
and \lya\ surface brightness in \har\ pixel-by-pixel.  We now discuss various
regions of the diagram:\\ 
$\bullet$ {\bf a}.  \lya\ can be emitted with fluxes similar to those expected
from recombination.  In Figure~\ref{fig:px2px} above log \lya\ surface
brightness of $\approx -12.8$, a region of proportionality is seen between \lya\
and \halpha\ that falls right on top of the expectation value for Case B.  These
pixels correspond to knot C in Figure~\ref{fig:h11}.  Since this region of
points is well defined and rather narrow, it can most easily be understood as
\lya\ photons leaving the galaxy with little interaction with surrounding \hI.
\lya\ photons emitted in these regions were most likely produced here. \\
$\bullet$ {\bf b}.  \lya\ can be emitted with fluxes below those expected from
recombination.  The preponderance of points in Figure~\ref{fig:px2px} lie below
the case B line.  This can be the result of two factors: \lya\ can be absorbed
by dust, decreasing the \lya/\halpha\ ratio (just as the Balmer decrement
increases with dust in nebulae), or \lya\ can be scattered out of the
line-of-sight by \hI.  I.e. dust and \hI\ scattering act to move
points down from the red line.  Recall, also, that in this logarithmic plot
\lya\ absorption cannot be visualized and more pixels are to be found at
negative values of \lya. \\ 
$\bullet$ {\bf c}. \lya\ can be emitted with fluxes above those expected from
recombination.  Toward the upper left region of the diagram, \lya/\halpha\
exceeds the value of 8.7 expected for Case B.  In this example, some pixels are
10 times brighter in \lya\ than expected.  This happens only at lower \halpha\
surface brightness, and here \lya\ is spatially redistributed from elsewhere and
emitted after scattering in the neutral ISM, resulting in the halo phenomenon
discussed above and in the following section.  Thus some of the \lya\ that shows
\lya/\halpha~$\lesssim$ Case B (point {\bf b}) must be scattered and not simply
attenuated by dust. \\

\subsection{Extended Halos}\label{sect:halos}

Early imaging observations showed that in galaxies with net \lya\ emission, the
dominant fraction comes from a component of large-scale extended emission that
surrounds the star-forming regions.  Indeed upwards of 50\% of the \lya\ is
typically emitted in halo regions that extended at least 10~kpc from the
UV-bright clusters \citep[][]{Hayes2005,Atek2008}, and in some galaxies this is
the only \lya\ that emerges.  The first large-scale \lya\ imaging survey of
local starbursts -- the \emph{Lyman alpha Reference Sample}
\citep[LARS,][]{Ostlin2014} -- has shown that extended halos are near
ubiquitous in \lya\ emitting galaxies \citep{Hayes2013}.  Moreover LARS has
enabled the first systematic survey of the sizes of these halos and the
comparison with other wavelengths.

\begin{figure}[t!]
\begin{center}
\includegraphics[width=0.9\textwidth, angle=0]{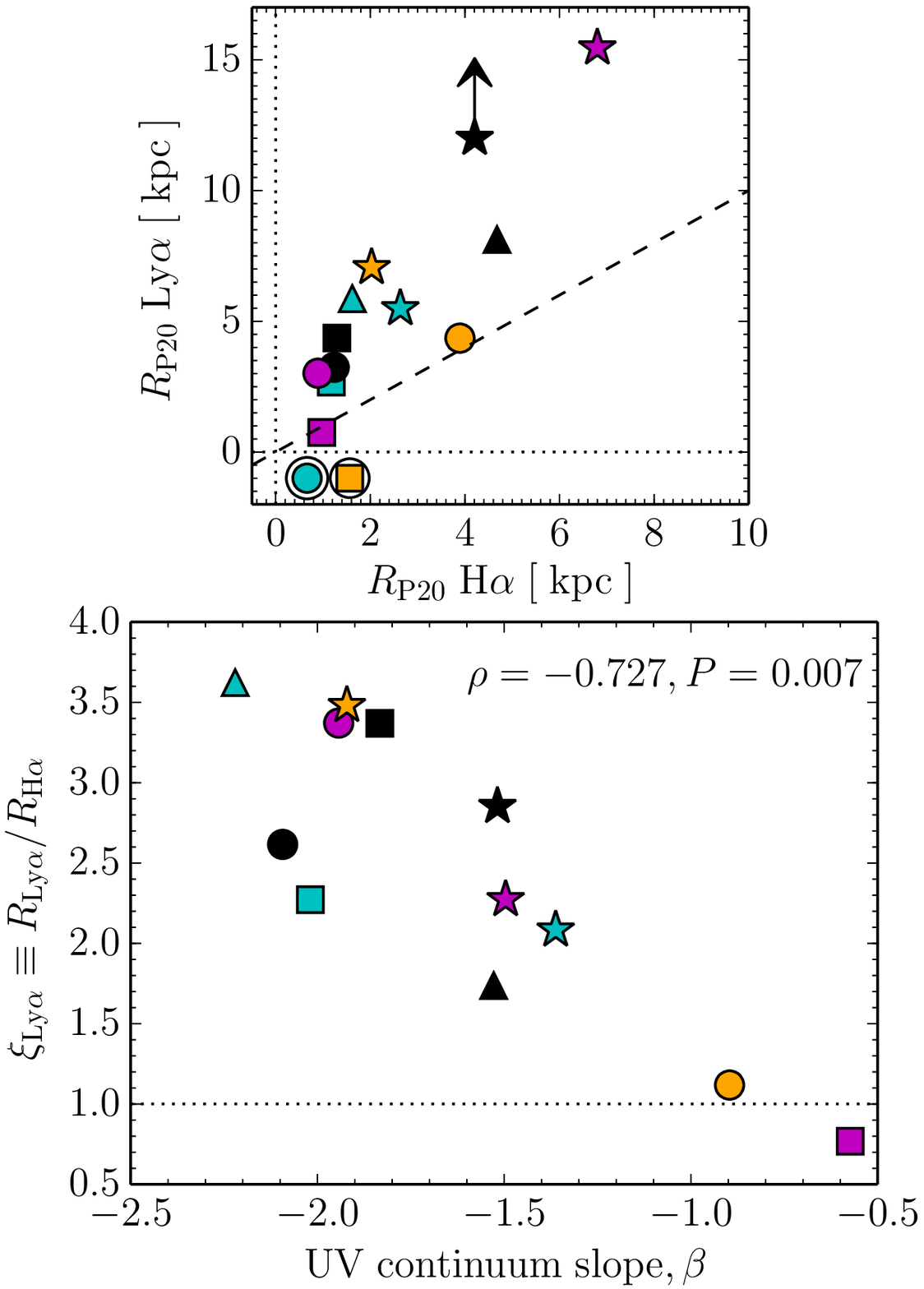}
\caption{Extension of \lya\ halos in the LARS sample; \emph{figure is taken from
\citet{Hayes2013}}.  The \emph{upper} panel shows the Petrosian radius measured
in \lya\ plotted against that measured in \halpha.  The two galaxies that
globally absorb \lya\ are set to negative size.  The dashed line shows the
1-to-1 line, and \lya\ is on average twice the linear size of \halpha.  The
\emph{lower} panel shows the relative extension of \lya\ compared to \halpha,
\xilya.  This example shows how the extension is anti-correlated with the UV
continuum slope, $\beta$, which is commonly used as a proxy for the dust content
in galaxies.}\label{fig:ext}
\end{center} 
\end{figure}

Continuum-subtracted \lya\ images show a wide range of morphologies.  Typically
they do qualitatively resemble those of the UV and \halpha, but are more
extended, and envelop the galaxies.  On average halos have twice the linear size
in \lya\ that the galaxy does in either the UV stellar continuum or \halpha\
(Figure~\ref{fig:ext}, \citealt{Hayes2013}).  These estimates are made using the
depth and redshift independent Petrosian radii, which are found to be below
15~kpc in \lya\ for LARS galaxies, with a median value of 5~kpc.  Furthermore,
the extension of the \lya\ surface (\xilya, the ratio of \lya\ radius to
\halpha\ radius) is not an independent quantity, and is correlated with a number
of measured properties: notably \xilya\ is anti-correlated with dust abundance,
as demonstrated by the lower panel of Figure~\ref{fig:ext}, and \lya\ extension
is found to be larger at lower metallicity and stellar mass.   Radiative
transfer simulations \citep[using][]{Verhamme2012} show that this effect cannot
be reproduced simply varying the dust content and what gives rise to these
extended halos is currently unclear.  \citet{Pardy2014} have shown that higher
\xilya\ is produced by galaxies with narrower 21~cm line-widths, but not the
total mass in \hI\ (which neglecting mergers correlates with the line-width),
possibly indicating that lower mass galaxies with less complex large-scale
morphologies are the ones in which \lya\ scatters to the largest relative
distances.

\section{LYMAN ALPHA SURVEYS AT LOW REDSHIFT}\label{sect:galex}

A common criticism of some local \lya\ studies is that the sample selection is
different from that of LAEs and LBGs at $z\gtrsim 2$.  The first generations of
small HST studies were assembled largely from IUE atlas \citep{Kinney1993},
which in turn were selected from older objective prism or BCG surveys; in
contrast most high-$z$ objects are selected by Lyman break techniques or \lya\
detection in narrowband filters.  Several recent studies have begun to rectify
this: firstly using HST the \emph{Lyman Break Analog} samples \citep[][based
upon UV luminosity and compactness]{Heckman2005,Hoopes2007}, the \emph{Lyman
alpha Reference Sample} (LARS; \citealt{Hayes2013,Hayes2014}, based upon UV
luminosity and \halpha\ EW), and studies undertaken with the GALEX Satellite
\citep[][based upon selection by \lya\ emission]{Deharveng2008,Cowie2010}. 

The GALEX satellite has been a vital contributer to low-$z$ \lya\ astrophysics.
As well as FUV and NUV imaging channels, GALEX also had the capability to
perform slitless spectroscopy across the same FUV and NUV bandpasses, providing
low-resolution spectroscopy of \lya\ for objects at $z\approx 0.19 - 0.45$ and
$z\approx 0.65 - 1.25$.  In turn, this enables us to really survey the low-$z$
universe for \lya-emitting galaxies \citep{Deharveng2008,Cowie2010}, and
redshifts around 1 \citep{Barger2012,Wold2014}, in a manner very similar to
those employed at high-$z$.  In the FUV channels GALEX LAE surveys
capture objects with NUV apparent magnitudes down to 21.8 (AB); at $z=0.3$ this
corresponds to a SFR of 3.6 \msunyr, assuming the continuum is unobscured.

\begin{figure}[t!]
\begin{center}
\includegraphics[width=0.9\textwidth, angle=0]{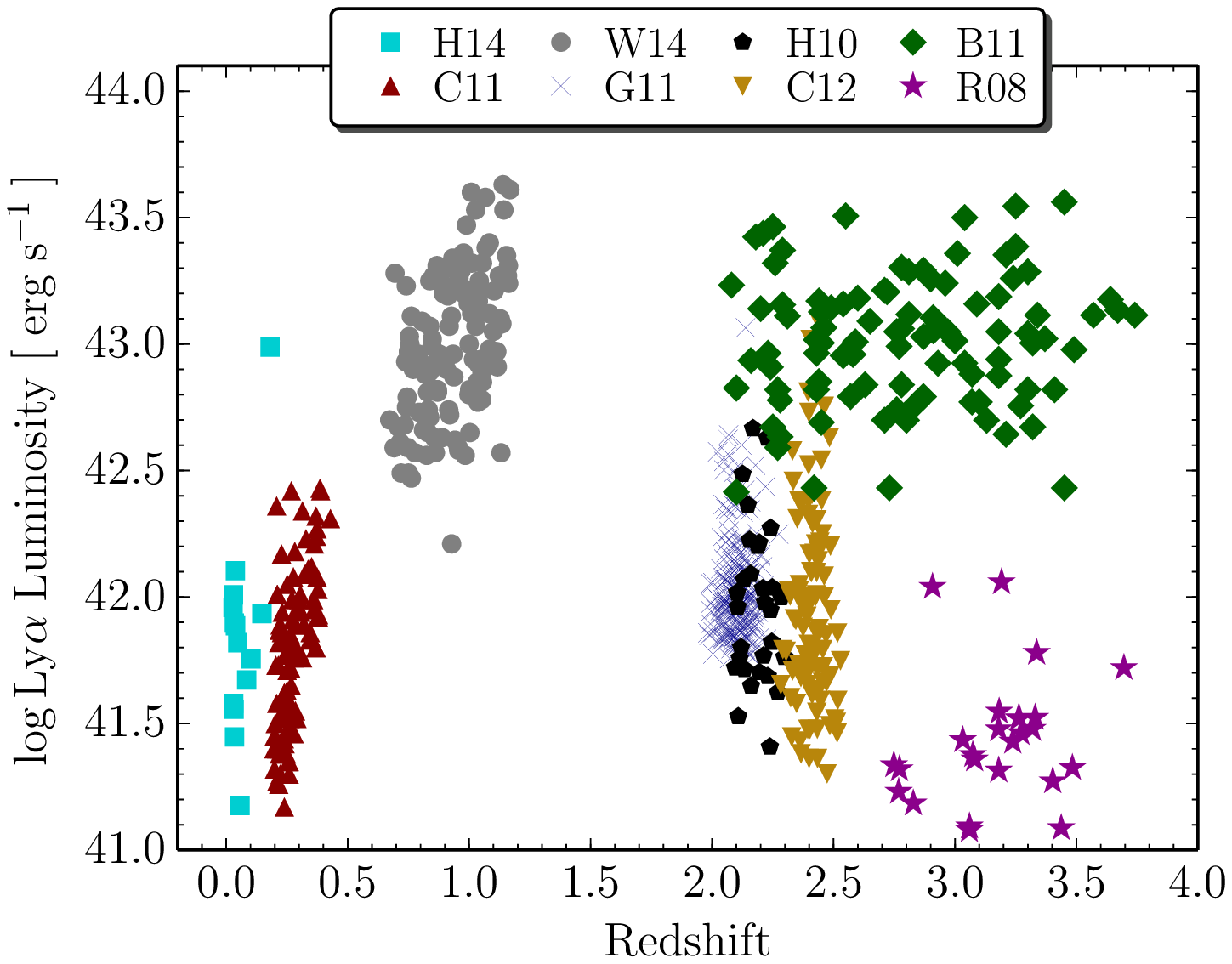}
\caption{The \lya\ luminosities probed by various galaxy surveys.  Reference
coding: C10=\citet{Cowie2011}; W14=\citet{Wold2014}; H14=\citet{Hayes2014};
R08=\citet{Rauch2008}; G11=\citet{Guaita2011}; B11=\citet{Blanc2011};
C12=\citet{Cantalupo2012}; H10=\citet{Hayes2010}.  Narrowband surveys (G11, C12,
H10) have been artificially randomized by $\Delta z=0.05$ to aid
visibility.}\label{fig:llya_vs_z}
\end{center} 
\end{figure}

Figure~\ref{fig:llya_vs_z} shows the \lya\ luminosities probed by various
studies out to $z\sim 4$.  As shown by the cyan squares and red upward-facing
triangles, UV-selected LARS galaxies and GALEX LAEs occupy a very similar range
of \lya\ luminosities.  Contrasting these luminosities with some $z>2$ surveys,
significant overlap is seen with the deepest ground-based observations: surveys
of \citet{Hayes2010}, \citet[][~both narrowband]{Cantalupo2012}, and
\citet[][blind long-slit spectroscopy, without aperture correction]{Rauch2008}
overlap the local HST and GALEX samples at above the 50\% level.

The $z\sim 1$ LAEs discovered by GALEX, however, are an order of magnitude more
luminous than the $z<0.4$ sample, and show almost no overlap with the more local
objects.  This is entirely a matter of distance, where the same limiting flux
corresponds to a factor of 20 difference luminosity between $z=0.3$ and 1.  This
difference in luminosity may introduce some biases when comparing the luminosity
distributions of the two GALEX samples, as we show in the coming Section.  These
$z\sim 1$ LAEs, however, do span a similar luminosity range to some of the
larger volume, shallower \lya-surveys at $z\gtrsim 3$ \citep[e.g.][]{Blanc2011}.

In the following Sections I discuss the most important results from the $\langle
z \rangle = 0.3$ and $\sim 1$ studies with GALEX, regarding the
numbers/evolution of \lya-emitting galaxies (Section \ref{sect:galaxevol}), and
statistical studies of their properties (Section \ref{sect:galexprop}); in part
of course the properties of galaxies are part of that evolutionary process. 

\subsection{Evolution of Lyman alpha Galaxies into the Nearby Universe}\label{sect:galaxevol}

\begin{figure*}[t!]
\begin{center}
\includegraphics[width=0.8\textwidth, angle=0]{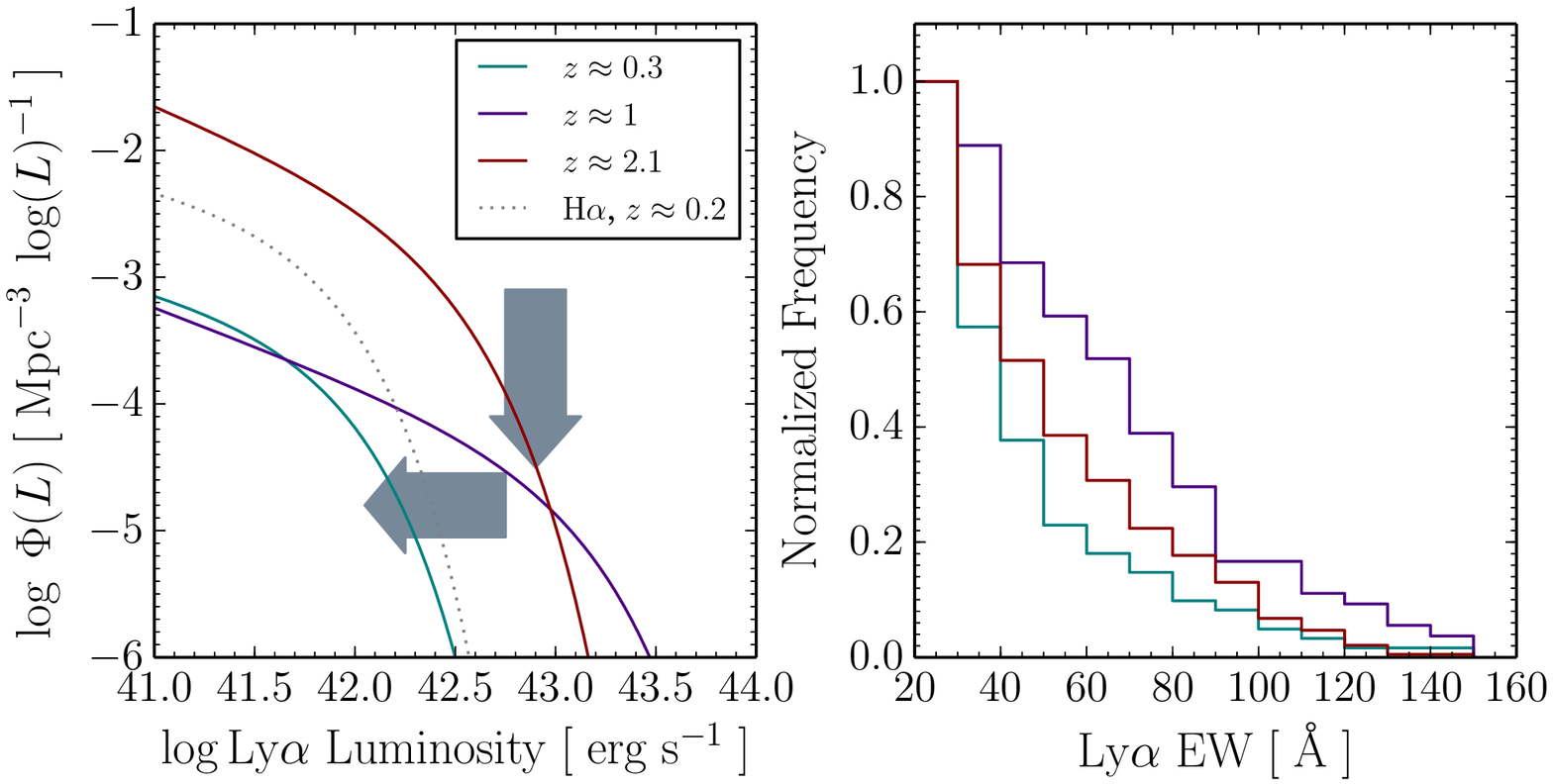}
\caption{\emph{Left}: The evolution of the \lya\ luminosity functions with
redshift.  Turquoise line shows the $z\approx 0.3$ LF of \citet{Cowie2010} and
purple shows the $z\approx 1$ LF of \citet{Wold2014}, measured by GALEX FUV and
NUV, respectively.  The dark red line shows the LF of \citet{Ciardullo2012},
measured at $z=2.1$ using narrowband filters.  Significant evolution is seen
between the distant universe and the local one; grey arrows show the suspected
modes of evolution in both luminosity and density (see text for details). The
dotted line shows the \halpha\ LF at $z=0.2$ of \citet{Tresse1998}.
\emph{Right}: the cumulative \lya\ equivalent width distribution at the same
redshifts; data are taken from the same surveys, with the $z\approx 2.1$
distributions from \citet{Guaita2010,Guaita2011}.}\label{fig:lfs}
\end{center} 
\end{figure*}

IUE demonstrated that \lya\ emission is rare in the nearby universe but GALEX
could determine how rare.  The first \lya\ luminosity functions (LFs) at $z
\approx 0.3$ showed that \lya\ emitting galaxies have become both fainter and
less abundant than at high-$z$.  Figure~\ref{fig:lfs} shows the LFs measured at
$z\sim 0.3$ and 1 \citep[][respectively]{Cowie2010,Wold2014}, together with
recent measurements for $z=2.1$ \citep{Ciardullo2012}.  

It should be noted that, while these LFs are the best that can be done with
GALEX, the samples are not large: 119 at $z\sim 0.3$ and 141 at $z\sim 1$.
However, because of GALEX's wide field-of-view and the large continuous
wavelength range provided by slitless spectroscopy, the cosmic volumes probed
are in fact rather large, and covering several $10^6$~comoving \mpccube.  Thus
while the individual parameters in the Schechter function may not be very
tightly constrained, the small number of galaxies is certainly because of the
relative paucity of LAEs at lower redshifts.

As discussed in Section~\ref{sect:galex}, the two GALEX samples cover different
luminosity ranges (Figure~\ref{fig:llya_vs_z}).  Moreover the dynamic range of
the surveys is not large, and even for the less luminous $z\lesssim0.4$ sample
it is not possible to calculate the faint-end-slope ($\alpha$) of the LF.  Thus
the LFs presented in Figure~\ref{fig:lfs} assume $\alpha$, basing the assumption
on measured values at $z=2-3$ for the $z\sim 1$ sample, and the $\alpha$ for
local \halpha-emitters for the $z\lesssim 0.4$ sample.  Note also that the
faint-end slope of the \lya\ LF is also not very tightly constrained at high
redshift \citep[reasons outlined in][]{Dressler2014}.

At $z\approx 0.3$ and to the UV limits mentioned in Section~\ref{sect:galex},
the shape of the \lya\ LF closely resembles that of \halpha\ and \hbeta,
although is lower in normalization: \lya-emitting galaxies make up about
$\approx 15$~\% of the local \halpha-selected galaxy population
\citep{Deharveng2008,Hu2009} and 1/20 the FUV counts at the same $z$
\citep{Cowie2010}.  This equates to a volume-averaged escape fraction of below
1~\% \citep{Hayes2011evol}, although this \fesclya\ will be slightly higher when
considering possible emission from galaxies not formally classed as LAEs (i.e.
galaxies that emit weaker \lya\, with $0\le W_{\mathrm{Ly}\alpha} < 20$\AA).  At
$z\sim 2$ the average \fesclya\ is $\approx 5$~\%, so the average \lya\ output
of the whole cosmic volume decreases 5-fold.  Note that this decrease in the
emitted fraction of \lya\ happens on top of the decrease in the cosmic star
formation rate density, which also drops by a factor of 5--10 over the same
change in redshift \citep{Madau2014}, implying the \lya\ luminosity density of
the local universe is far below that of $z\gtrsim 2$.

A large fraction of this evolution takes place in the 4.3~Gyr that elapses
between redshifts of 1 and 0.3 \citep{Wold2014}.  At the higher redshift of 1
luminous \lya\ emitters are certainly in place, with luminosities equivalent to
\lstar\ at $z\approx 2-3$,  and the evolution between $z\approx 0.3$ and 1 can
be well described by a simple factor of 10 increase of \lstar\ in the
\citet{Schechter1976} function.  Over this redshift range the space density of
LAEs does not appear to change but the galaxies simply scale up in luminosity.
Between $z=1$ and 2 (3 Gyr), $\phi^\star$ increases by an order of magnitude to
produce the LFs observed in ground-based surveys.  Arrows in
Figure~\ref{fig:lfs} show how the evolution of the LF manifests as a drop in
density, followed by a dimming.  Between the peak in the cosmic SFRD at $z\sim
2-3$ and $z=1$, the universe first acts to turn off a fraction of the \lya\
emitters, whereafter between $z=1$ and the nearby universe the abundance is
constant but the galaxies get fainter in line with the evolution in both the UV
and \halpha.  

Interestingly, this evolution is not strongly reflected in the equivalent width
distribution of LAEs, which does not evolve as dramatically.  The EW
distribution of \citet{Guaita2011} at $z\approx 2.1$ agrees well with the
distribution at $z\sim 1$ (shown in the \emph{right} panel of
Figure~\ref{fig:lfs}) despite the fact that the overall \lya-emitting fraction
has decreased by a factor near 5 (\citealt{Cowie2010} contrasted with
\citealt{Shapley2003}).  Thus whatever process is turning \lya-emitters off it
does not affect the shape of the remaining EW distribution (i.e. galaxies with
\ewlya~$\gtrsim 20$\AA).  Of course the equivalent width distribution of the
overall population changes significantly, as many galaxies drop below the
canonical 20\AA\ limit; the higher EW tail of the distribution remains largely
constant.  Note that it is not necessarily fair to conclude strong evolution in
the EW distribution to $z\sim 0.3$ from Figure~\ref{fig:lfs}, as the GALEX FUV
observations are first continuum selected. This will lead to a fraction of
continuum-faint objects with high EWs being missed, which will extend the tail
of the distribution.  However because the bulk of the luminosity comes from low
EW galaxies, the LFs are will be largely unaffected.

In light of the extended \lya\ halos discussed in Section~\ref{sect:halos}, we
may ask whether much \lya\ also extends beyond the spectroscopic extraction
apertures of GALEX.  These spectra are extracted using an optimal model of the
point spread function \citep[PSF,][]{Morrissey2007}, which has FWHM of $\approx
5$\arcsec.  At $z=0.3$ ($z=1$) this aperture corresponds to a spatial scale of
22 (40) kpc, and thus one dimensional spectra are summed over scales that exceed
this.  At $z\sim 0.3$ the aperture is 5 times the median \lya\ Petrosian radius
in LARS, and while we do not known precisely how the \lya\ profiles behave at
larger radii, it is likely that in most cases GALEX captures the majority of the
total flux.  At $z=1$ \citet{Barger2012} discovered a giant \lya\ blob, extended
over an 18\arcsec\ diameter, but also measure such objects to be extremely rare
(one in the whole volume).  Moreover, aperture sizes ($\approx 5$\arcsec\ at
$z=0.3$) are equivalent to a 3\arcsec\ diameter aperture at $z=2$, so if halo
extension does significantly affect the recovered \lya\ flux, it is likely by a
similar factor as in high-$z$ observations.

\subsection{The Properties of Nearby Lyman alpha Galaxies}\label{sect:galexprop}

\begin{figure*}[t!]
\begin{center}
\includegraphics[width=0.9\textwidth, angle=0]{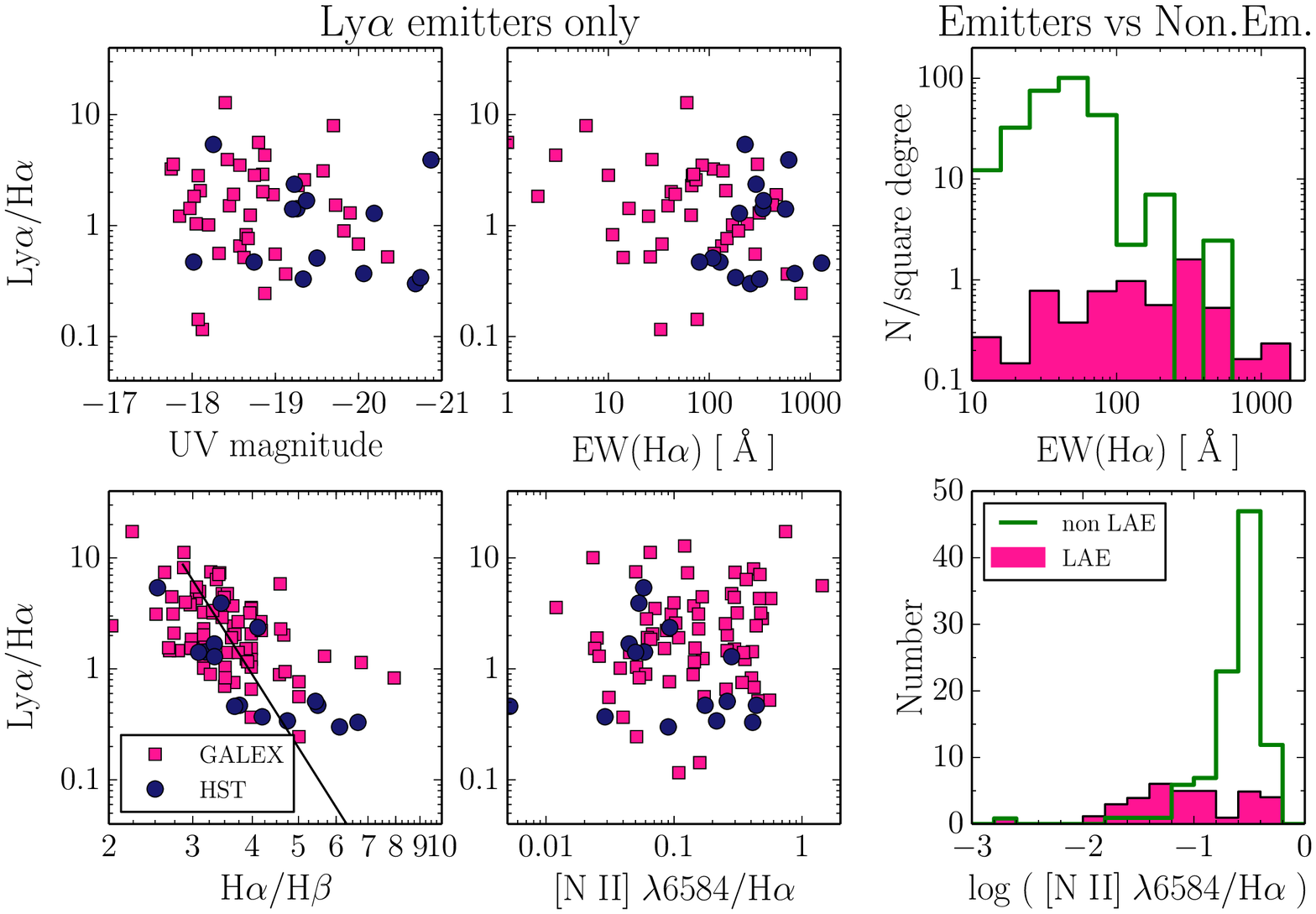}
\caption{Scatter plots to the \emph{Left} show the observed global \lya/\halpha\
ratio measured by GALEX (pink squares) and HST (dark blue circles).  For the
GALEX samples, data are assembled from \citet{Cowie2011}, \citet{Scarlata2009},
\citet{Finkelstein2011galex}, and \cite{Atek2014galex}; for the HST samples data
are taken from \citet{Ostlin2009,Ostlin2014} and \citet{Hayes2013,Hayes2014}.
Note that because of the differing quantities plotted on the abscissa, the
figures do not necessarily include the same number of points.  Furthermore,
since these points are chosen to have \lya\ in emission, the non-emitting
population cannot be visualized. \emph{Upper Left}: the absolute UV magnitude in
the AB system.  \emph{Upper Right:} the \halpha\ equivalent width.  \emph{Lower
Left}: the \halpha/\hbeta\ ratio, where the black line shows the effect of dust
attenuation (\citealt{Calzetti2000} law), assuming \lya/\halpha=8.7 and
\halpha/\hbeta=2.86.  \emph{Lower Right}: the [\nII]$\lambda 6584$\AA/\halpha\
ratio (=N2 index).  Histograms to the \emph{Right} compare the frequency of
GALEX-LAEs (pink filled histograms) and non-\lya-emitting galaxies (green
outlined histogram), as a function of \ewha\ (\emph{upper}) and N2 index
(\emph{lower}).}\label{fig:globtrends}
\end{center} 
\end{figure*}

Unlike most high-$z$ studies, both the GALEX and HST samples are sufficiently
close that many of their properties may be systematically measured.  I assemble
some of the key data obtained from these telescopes in
Figure~\ref{fig:globtrends}, and in this Section discuss what we have learned
about galaxies that emit, and do not emit, \lya.

\subsubsection{Stellar Populations}\label{sect:galexprop:stelprop} 

Over the \lya\ luminosity ranges probed by low-$z$ \lya\ observations ($\sim
10^{42}$~\ergsec), known LAEs have typical stellar masses on the order of $\sim
10^{9}$~\msun\ \citep{Finkelstein2011galex,Hayes2014}.  This is significantly
higher than estimates made for most LAE samples at high-$z$, while the \lya\
luminosities are equivalent or somewhat lower.  This difference in the
$M/L$(\lya) reflects the fact that the average nearby galaxy is more evolved,
and has had substantially more time to build stellar mass than a similarly
selected galaxy in the more distant universe.  

This evolution is also reflected in recovered stellar ages. However caution is
needed here: when estimated from SED fitting age is a luminosity-weighted
average, and wavelength-dependent function of the included bands and the range
of permitted star-formation histories.  \cite{Finkelstein2011galex} find a wide
range of SED-fit ages with a median of $\sim 300$~Myr.  However when subtracting
an underlying population of old stars, \citet{Hayes2014} determine recent
star-formation to have occurred in short bursts with luminosity-weighted ages up
to a few tens of Myr.  This suggests that \lya\ emitting galaxies, like
starbursts in general, have temp temporarily elevated SFRs.  The total \lya\
luminosity shows no dependency upon stellar age, but both \ewlya\ and the
relative throughput (\fesclya) are higher in younger galaxies. 

The above would suggest that \lya\ escape fractions and equivalent widths are
higher in galaxies with higher \halpha\ equivalent width, which roughly measures
the specific SFR (sSFR).  \citet{Cowie2011} indeed show that the fraction of
\lya-emitting galaxies increases when progressively higher \halpha\ EW
thresholds are applied, finding almost 60\% LAEs in sub-samples of
\ewha$>250$\AA.  \citet{Hayes2014} similarly find that all their galaxies with
\fesclya\ above 10\% have \halpha\ EWs above 400\AA.  Interestingly, however
this effect is not visible when comparing the \lya/\halpha\ ratio (or \ewlya)
with \ewha\ (Figure~\ref{fig:globtrends}) for the objects with \ewlya$>20$~\AA.
The most convincing results emerge when we compare the average properties of LAE
and non-LAE samples.

\subsubsection{Galaxy Morphology}\label{sect:morph}

In UV-selected local galaxies, the objects with the highest \ewlya\ and
\fesclya\ are found to be among the more compact ones.  LARS galaxies tend to
have particularly compact UV morphologies and LAEs are found among those with
Petrosian radii of $\sim 1$~kpc on average, similar to higher redshift results
of \citet{Malhotra2012}.  Larger galaxies all show lower \ewlya.

Galaxies hosting \lya-emitting starbursts represent a mixture of various types,
although still a mixture that is distinct from the UV-selected galaxy population
in general.  The LAEs contain a higher fraction of compact galaxies and merging
systems.  More curiously, an enhanced fraction of the GALEX LAE disks appear to
be face on \citep{Cowie2010}.  This phenomenon is also expected from radiative
transfer modeling \citep{Verhamme2012,Laursen2013}, and suggests that
orientation effects may hide some disks from \lya\ selection.  Unfortunately
current samples are not sufficiently large to test \lya\ emission as a function
of inclination angle.

\subsubsection{Interstellar Dust and Metals}

\noindent
\paragraph{Dust}
\mbox{}\\

LAEs have bluer UV-optical colours than non-emitting galaxies of the same
magnitude \citep{Cowie2010,Cowie2011} but there is significant overlap in
colours between the emitting and non-emitting subsamples.  UV-selected samples
similarly exhibit higher \lya\ escape fractions (\fesclya$>10$~\%) where UV
colours are bluer ($\beta < -1.8$; \citealt{Hayes2014}).  This should be in part
a reflection of the stellar age effects discussed in
Section~\ref{sect:galexprop:stelprop}, but may also be due to dust extinction
that reduces the \lya\ throughput: \lya/\halpha\ ratios and escape fractions
(independent of age) are both higher for bluer galaxies, not only equivalent
widths.

The average \lya/\halpha\ ratio found for the GALEX-selected LAEs is slightly
above 2 \citep{Atek2009galex,Scarlata2009,Cowie2010}, although perhaps
surprisingly, plots comparing \lya\ and \halpha\ flux directly show no general
covariance over more than 1~dex in each quantity.  Examining \lya/\halpha\
however, the ratio decreases significantly with increasing \halpha/\hbeta, and
the conclusion that dust reduces the transmitted \lya\ in the LAE samples is
shared over many studies.  Indeed as shown in the \emph{lower left} panel of
Figure~\ref{fig:globtrends}a, the anticorrelation between \lya/\halpha\ and
\halpha/\hbeta\ is one of the few trends that is significant over the dynamic
range of today's surveys.

The comparison of these line ratios with extinction laws reveals several curious
features.  Firstly, at the lowest \halpha/\hbeta\ ratios, most of the galaxies
lie below the predicted curves, and even in \lya-selected samples only $\sim
25$\% of \lya\ photons escape when \halpha/\hbeta\ is in the range 2.8--3.2.
These galaxies emit less \lya\ than dust attenuation would predict.  However the
same is not true for dustier galaxies: at \halpha/\hbeta~$\gtrsim4$ the mean
\lya/\halpha\ ratio is $\sim 1$, even though from this Balmer decrement we would
would expect 97\% of the \lya\ radiation to be absorbed.  Thus the locus of
points in the \lya/\halpha--\halpha/\hbeta\ plane shows a trend that is much
flatter than known extinction laws, and the \emph{normalized} \lya\ escape
fraction (measured \fesclya\ divided by that which is expected for the derived
dust content) increases with attenuation.  Above \ebv~$\approx 0.3$ \lya\
emission becomes on average several times stronger than expected
\citep{Scarlata2009,Atek2014galex,Hayes2014}.

The most basic example of a simple screen of dust that reddens the nebular lines
is incompatible with observation.  \citet{Atek2009galex} and
\citet{Finkelstein2009galexpops} invoked the \citet{Neufeld1991} geometry to
explain this apparent enhancement of \lya, in which dust is embedded within the
\hI\ clumps of a multiphase ISM, and \lya\ scattering prevents photons from
encountering dust.  Radiative transport simulations show that this effective
`boost' of \lya\ is very difficult to reproduce without rather contrived
combinations of parameters \citep{Laursen2013,Duval2014}, and the predicted
increase of \lya\ EW with measured attenuation is not observed.

\citet{Scarlata2009} argue instead for a scenario that requires no such
preservational scattering, but is still built upon a clumpy dust distribution
(which may anyway follow the cold gas).  This model does not require clumps to
act as mirrors to \lya\ and nor does it predict \lya\ EW to rise with \ebv.
Similarly \citet{Atek2014galex} point out that a galaxy is likely made up of
many \hII\ regions with a large variety of optical depths and since we see only
down to an optical depth of 1 at each wavelength, observed radiation in each
line (and continuum) comes from regions of different sizes.  At the dusty end of
the galaxy distribution these latter two scenarios do not require any scattering
at all, and indeed recently \citet{Martin2015} have shown that significant \lya\
emission can be detected from ultraluminous infrared galaxies (ULIRGs), from
which it most likely escapes through holes in the ISM.

\paragraph{Metal Abundance}
\mbox{ }\\

Dust can absorb \lya\ and metals cannot.  At the UV luminosities probed by GALEX
and LARS, the LAEs exhibit metallicities that overlap with the UV-continuum
selected galaxies, but extend down to lower metal abundance.  LAEs are on
average deficient in nebular oxygen by about 0.4~dex
(Figure~\ref{fig:globtrends}c), and are drawn mainly from a sub-populations with
metallicities of $12+\log(\mathrm{O/H}) \lesssim 8.2$.  Indeed a remarkable
result from the GALEX studies is that the N2 index (=log([\nII]/\halpha)
segregates \lya-emitters from non-emitters more cleanly than \halpha/\hbeta. 

Part of this apparent preference for low metallicity galaxies comes from an
enhanced fraction of more compact irregular galaxies (Section~\ref{sect:morph}),
which are lower metallicity in general.  Comparing \lya\ EW with age and
metallicity, \citet{Cowie2011} suggest LAEs are drawn from a stage in the
evolutionary sequence during which metals build up and \lya\ EWs decrease, in a
similar way to that which would be expected for \halpha.  In addition
\citet{Hayes2014} find that not only does EW decrease with age but also the
\lya\ escape fraction, which has no dependence upon stellar evolutionary stage.
This shows that not only do older stellar populations produce less \lya, but
also that their ISM become more opaque to \lya; this would also support
hypothesis of dust buildup, although current samples do not have the statistical
power to say whether this is purely a dust effect.

\subsubsection{Neutral Gas} 

While dust absorption is ultimately the process that can expunge \lya\ radiation
from a galaxy, \hI\ scattering determines the path length of \lya\ between the
nebulae in which it is produced and eventual emission.  Here we discuss
observations that probe the \hI\ phase directly by 21~cm emission, telling us
about the total amount of \hI\ available for scattering, and its large scale
kinematics.  For a discussion of the \hI\ properties measured on small scales by
absorption line studies, see Section~\ref{sect:hstspec}. 

Masses in \hI, and the large scale \hI\ envelope of individual galaxies can only
be measured by 21~cm observations, which currently limits us to the nearby
universe.  However to avoid \lya\ absorption by the Milky Way and to separate
the line from geocoronal emission, the lower limit of a galaxy's recession
velocity is $cz\lesssim 2500$~\kms.  By this distance of $\approx 35$~Mpc, \hI\
21~cm observations are already challenging \citep{Pardy2014}. 			
 
Regarding the \lya-absorbing dwarf galaxies, \izw\ is known to lie within a huge
\hI\ envelope that is many times its optical size, with a central column density
of \nhi~$\gtrsim10^{21}$~\percm\ \citep{vanZee1998} and total mass of $7\times
10^{7}$\msun.  Thus if all the \hI\ were static with respect to the \hII\
regions, \lya\ would see upwards of $10^7$ optical depths at line centre.  \sbs\
exhibits very similar properties in both central \lya\ absorption and large
static \hI\ envelope \citep{Thuan1999}.  In cases like this the fate of the
\lya\ radiation is unclear: does the \lya\ scatter so many times that it is
eventually absorbed by the small amount of available dust, or does the \lya\
eventually leak out in a large, low surface-brightness halo that cannot easily
be detected?  Without very large aperture imaging observations that cannot be
obtained with existing facilities, conclusive answers are difficult to provide.

Sixteen more starbursts ($M_\mathrm{UV} = -17.5$ to $-21$, from
\citealt{Ostlin2009} and the LARS sample), complete with \lya\ imaging and
spectroscopy from HST, have been observed at $\lambda=21$~cm with the Green Bank
Telescope (GBT) and/or Karl G.  Jansky Very Large Array
\citep[VLA,][]{Cannon2004,Pardy2014}.   \hI\ masses are between $10^9$ to
several $10^{10}$~\msun, which is or 10--100 times as much as the dwarf galaxies
, but for the seven galaxies where \hI\ detections are resolved the central
column densities are comparable.  In these samples the \lya-emitting objects
with \fesclya\ above 10\% all have \hI\ masses below $4\times 10^{9}$\msun.
While galaxies above this mass emit only a small fraction of their \lya\
photons, several of the less \hI-massive galaxies also show \fesclya\ below a
few percent, or in absorption.  The same result is seen with dynamical masses
derived from \hI\ line-widths.  While it may be hypothesized that less \hI\
would permit more \lya\ photons to escape, it cannot be ruled out that this
anti-correlation between \fesclya\ and \mhi\ is not simply reflecting the fact
that higher \fesclya\ is also seen at lower stellar mass
(Section~\ref{sect:galexprop:stelprop}) and is not causally connected with 
\hI.

Spatial resolution in these 21~cm observations is currently still low, even with
the VLA, but in every case the \hI\ is resolved, and extends over more than 10
Petrosian radii in the UV.  Most galaxies show signs of tidal interactions, both
in their 21~cm morphologies and line shapes.  Perhaps most interestingly \eso,
\iras\ \citep{Cannon2004} and Arp\,238 \citep[LARS\,03,][]{Pardy2014} show long
\hI\ tidal streams and debris trails: in the cases of \eso\ and \iras\  these
streams stretch over 50--100~kpc to companion objects, while in Arp\,238 a huge
($M\sim 2\times 10^{9}M_\odot$) \hI\ body has been ejected that contains no
apparent companion galaxy at the level of SDSS imaging.

\section{SYNTHESIZING THE OBSERVATIONAL DATA: WHAT MAKES A LYMAN ALPHA EMITTER?}\label{sect:govprop}

As we have seen in Sections~\ref{sect:hstspec}, \ref{sect:hstimage}, and
\ref{sect:galexprop}, current data show that \ewlya\ and \fesclya\ are
influenced by a large number of physical properties.  In recent years much
attention has been devoted to determining their order of precedence: i.e.
whether covering fraction is more important than kinematic properties, which in
turn is more important than dust reddening.  Mostly we have been driven to find
a way to predict the emergent \lya\ flux, EW or \fesclya\ from a given set of
conditions.  We have now assembled significant samples of low-$z$ observations,
selected by \lya, \halpha, and UV continuum, and have studied various subsets of
them between the X-ray and radio, measuring all the physical properties
discussed in Section~\ref{sect:galex}.  By combining this literature we now have
substantial power to determine how \lya\ is regulated and test for the primary
effects in homogeneously selected galaxies that do and do not emit \lya. 

Simultaneously we need to explain:
\begin{itemize}
\item{Why only $\approx 5$\% of local UV-selected galaxies (down to the SFRs of
$\approx 4$\msunyr, that are within reach of GALEX) show \ewlya\ above 20~\AA.}
\item{Why \fesclya\ and \ebv\ are anti-correlated in \lya-selected galaxies, but
many galaxies significantly outlie this relationship; that many seemingly
dust-free galaxies absorb their \lya, while many dusty galaxies exhibit
\fesclya\ that appears too high for their extinction.}
\item{Why higher \ewlya\ and \fesclya\ are found among galaxies with lower
stellar mass and metallicity, and younger stellar age.}
\item{Why higher \ewlya\ and \fesclya\ are found among more compact galaxies and
face-on spirals.}
\item{Why \lya\ emission, at least on small scales, is frequently associated
with galaxy outflows, and the spectroscopic line profile is almost always
asymmetric.}
\item{Why among our highest EW \lya-emitting galaxies we often infer covering
fractions below unity (measured from, \cII\ and \siII) or low \hI\ column
densities.}
\end{itemize}

\noindent 
In doing so we must remain cognizant of the fact that data have been assembled
from different selection functions, measurements have been made in apertures
that probe a variety of physical sizes, and that large scale halo emission may
affect some, but not all, of our results. 

We first address stellar properties, as these dictate the intrinsic \ewlya\ but
do not affect transmission (i.e. they have no direct influence over \fesclya,
and only set a limit on the emergent \ewlya).  Since \lya\ is reprocessed
ionizing continuum, \ewlya\ will exceed 20\AA\ only during for ages below
$\approx 6$~Myr \citep[][and
Section~\ref{sect:observable:stars}]{Leitherer1999}, assuming an SSP.  The SSP
assumption is probably over-simplistic for whole galaxies but it does serve to
illustrate that the \lya-bright period cannot be sustained if the SFR is
declining, and that episodes must be young.  The \lya\ production must be most
strongly correlated with the evolutionary stage of the stars. 

However the observed \ewlya\ is not strongly correlated with instantaneous SFR,
and neither are the high-EW galaxies the most luminous in \lya.  If anything the
reverse is true, at least within the current samples, and over the UV and \lya\
luminosity ranges probed at low-$z$ high-EW galaxies are among the less
luminous.  Indeed similar can be said for galaxy samples selected by other
emission lines such as \halpha\ or [\oIII]$\lambda 5007$
\citep[e.g][]{Atek2011}: the requirement for high EW line emission necessitates
not only ongoing star formation, but at a given SFR the UV continuum must be
faint enough for the EW to be high.  This at least partially explains why local
LAEs are drawn from galaxies with higher \halpha\ EWs \citep[as seen by,
e.g.][]{Cowie2011}, and at lower stellar age and mass.  Considering the
well-known mass-metallicity relationship \citep{Tremonti2004}, it equally well
explains why LAEs should be more prevalent at lower metallicities.  Note,
however, that according to current spectroscopic data, this relationship does
not extend indefinitely to the lowest metallicity, gas-rich dwarf galaxies; a
point to which we will return below.

\lya\ \emph{transfer} however is not affected by any of the above
considerations: stellar age, mass, and nebular oxygen abundance are not
properties that influence radiation, although they may correlate with quantities
that do.  What matters from this point is the properties of dust and \hI.
\fesclya\ is anticorrelated with the \ebv, as would be expected, but the spread
is very large.  Moreover the slope of the \fesclya--\ebv\ relationship in LAE
samples is flatter than would be expected for pure extinction
\citep{Atek2014galex}, indicating that the role of dust is diminished.  At the
dust-free end the under-luminous \lya\ can be most easily be explained by the
presence of \hI, as it may scatter \lya\ photons many times, and increase the
probability of dust absorption.  This is supported by, for example, small
aperture spectroscopic observations of almost dust-free dwarf galaxies that show
not only a lack of nebular \lya\ but also the absorption of continuum photons in
the \lya\ resonance.

If this dust were distributed purely as a screen surrounding the star-forming
regions, the apparent over-luminous \lya\ observed in some LAEs cannot be
explained without severely modifying extinction laws.  \lya\ would have to see
at least the expected attenuation, plus an excess of absorption because of
scattering.  To explain this we must invoke geometrical effects.  Evenly mixing
dust into the \hII\ regions would produce an effective attenuation of the form
\fesclya~$\propto 1/\tau_\mathrm{dust}$ at high $\tau_\mathrm{dust}$, and
asymptotically sets a lower limit to \lya/\halpha\ ratio of $\approx 2$ (without
scattering).  However such a dust geometry cannot make high a \halpha/\hbeta\
ratio, which saturates for optically thick nebulae at $\approx 4$.  What can
explain the simultaneous high \lya/\halpha\ and \halpha/\hbeta\ ratios is a
clumpy distribution of dust, as advocated by \citet{Scarlata2009}.  This model,
originally implemented by \citet{Natta1984} and \citet{Caplan1986}, assumes dust
to be distributed in dense clumps and the effective attenuation law changes
significantly with the average number of clumps along the line-of-sight.  E.g.
again without scattering, varying $\tau_\mathrm{dust}$ within 10 clumps may
produce \lya/\halpha$>2$ with \halpha/\hbeta$\approx 7$, and can explain the
line ratios seen in the dustiest local LAEs.  The fact that \lya\ is seen at all
from local ULIRGs \citep{Martin2015} also suggests that \lya\ must find paths of
low dust optical depth.

Then we need to explain why the local dwarf galaxies (e.g. \izw, \sbs) centrally
absorb at \lya\ despite their very low dust contents.  Indeed HST spectroscopy
shows some absorption component in every object observed, demonstrating that
\hI\ has effects that range between small dips and very broad, damped absorption
features.  \citet{Tenorio-Tagle1999} presented an evolutionary sequence for the
\lya\ spectral profile expected from a star cluster (assuming an SSP).  This
model assumes that during the earliest stages of a cluster's evolution
($\lesssim 2$~Myr) local \lya\ absorption would be expected because the
surrounding medium is completely static.  Over the subsequent $\sim 4$~Myr
mechanical feedback from O star winds and the first supernovae accelerate the
\hI\ outwards, producing \pcyg-shaped \lya\ profiles.
The application of this model to dwarf galaxies suggests that their
star-formation episodes may be too young and that feedback has not yet had time
to accelerate the neutral ISM \citep{Mas-Hesse2003}.

This scenario is probably an over-simplistic representation of whole galaxies,
where stars form in different regions over extended timescales, but nevertheless
the qualitative arguments may be helpful in understanding the influence of
galaxy winds.  All the observed line profiles, including those of the local
dwarf galaxies, can in principle be unified within such a scenario
\citep{Mas-Hesse2003}.  In the complete literature of low-$z$ galaxies only one
example shows symmetric \lya\ emission without an obvious absorption component
\citep[Tol\,1214--277,][]{Thuan1997}, and as Figure~\ref{fig:ewdv} shows, an
outflow in the neutral medium appears to be a requirement (but not uniquely
sufficient) for net \lya\ emission
\citep{Kunth1998,Wofford2013,Rivera-Thorsen2015}.

\section{PERSPECTIVES AND FUTURE DEVELOPMENTS: THE MOST PRESSING
QUESTIONS}\label{sect:conc}

With GALEX and HST \lya\ observations now running to over hundred galaxies for
each telescope, local samples are substantial.  Nevertheless, several major
questions still remain to be answered.  This closing Section is dedicated to
outlining a handful of the most pressing questions that, while perhaps
challenging, can be addressed within current samples, or with possible
extensions with present-day facilities.

\subsection{What is the Atomic Gas Distribution in Star-Forming Galaxies?}

The spatial distribution of emitted and absorbed \lya\ must reflect some set of
properties of the \hI\ gas.  In very nearby galaxies, 21~cm observations have
already revealed the structure of the cold atomic gas where, for example, the
\emph{\hI\ Nearby Galaxy Survey} (THINGS, \citealt{Walter2008}) and VLA-ANGST
\citep{Ott2012} surveys have found an atomic medium that is largely
inhomogeneous and clumpy.  It is clear that if \lya\ photons were injected in
these galaxies they would experience more scattering in some regions than
others.  Examining at least these 21~cm observations, we see that the \hI\ does
not much resemble the shells and slabs in which many radiation transfer
calculations are done.  In order to empirically determine the effects that \hI\
and its distribution have on \lya\ emission requires resolved observations of
individual targets in both \hI\ and \lya. 

Currently we may contrast average \lya\ surface brightness profiles
\citep[e.g.][]{Hayes2014} with those of the intensity at 21~cm
\citep[e.g.][]{Bigiel2012}.  These samples show both \lya\ and 21~cm intensity
profiles that are best fit with S\'ersic profiles that are close to exponential
($n\sim1$), while UV/optical wavelengths in the LARS sample show significantly
higher S\'ersic indices ($n\gtrsim 3$).  Does the \lya\ surface brightness trace
the gas column density?  While curious, this may be entirely coincidental as the
sample selection is very different in the two cases.  Such observations
established in the same galaxies would be enormously instructive in interpreting
the \lya\ halos of both galaxies at low and high redshifts.

It is unfortunate that resolutions attainable in 21~cm and in the UV/optical are
not well matched; sampling small physical scales at 21~cm requires very local
galaxies, while the redshift requirements to observe \lya\ imply targets must
lie beyond several tens of Mpc.  Progress has been made by \citet{Cannon2004}
and \citet{Pardy2014}, but with synthesized beam sizes of $\approx 15$~arcsec in
the best case (usually much coarser), such \hI\ observations can place only two
resolution elements inside the linear size of the ACS/SBC camera.  However VLA
in configurations B and A can provide resolutions down to around 4 and 1.3
arcsec, respectively, albeit with a substantial increase in observing time.
Observational programmes at higher resolution (VLA C and B configurations) are
ongoing, but it is clear that large steps forward may be taken if \hI\
observations are pushed to the highest spatial resolutions, and such
observations are now essential.

\subsection{How is the \lya\ Spectral Profile Built?}

The total integrated \lya\ profile of a galaxy is built by emission from
different regions, likely with differing kinematics, orientation, and with
different contributions to the total \lya.  The spectral profiles of \lya\
measured in small apertures tend to be \pcyg, with radiation absorbed from the
blue and re-radiated in the red.  In these small apertures, \lya\ escape
fractions are also low.  However imaging tells us that these photons are often
not expunged from the system, but scattered back into the line of sight at
different position, where large-aperture photometry measures significantly
higher \fesclya.  It is an unfortunately common feature of absorption
spectroscopy that we rarely know the line-of-sight distance between the emitting
sources and the absorbing gas. 

Ideally we would like to know where the frequency redistribution of \lya\
occurs, and whether the halo emission shares a similar profile to that of the
more central regions where the starburst is located.  For example if frequency
redistribution occurs close to ionizing clusters and \lya\ is singly scattered
at large radii, the profile may indeed be similar over large distances.  However
if photons also get caught for many scatterings in halo gas then a new kinematic
structure may be encoded in the \lya, or double-peaked profiles may be seen.

To resolve this we may take high resolution spectra of the diffuse \lya-emitting
regions in galaxy halos.  These observations would be best supported by
aperture-matched optical spectroscopy of \halpha\ or \hbeta, to tightly
constrain the rest velocity distribution of \lya, and ideally also the highest
possible resolution 21~cm observations (previous section).  The best tool for
this would be HST/STIS with narrow slits, which could provide resolutions of
around 20~\kms, but again at the cost of long integrations in faint regions.
\citet{Mas-Hesse2003} have shown that in \iras, the blue wing of the \lya\
profile sets on at a similar wavelength over at least 10~kpc, suggesting the
star forming regions are surrounded by a \hI\ medium that may be rather
homogeneous in both space and velocity.  However 21~cm observations find gas at
much larger radii, where \lya\ has not yet been spectroscopically detected;
whether the profile bends smoothly and traces the edge of a bubble, and how this
effect may vary in different galaxies are all currently unknown.

\subsection{Ionization State of the Interstellar Medium}

Section~\ref{sect:galexprop} discusses the effect of a large number of galaxy
properties on the emission of \lya.  Many have been studied over time, but
largely overlooked has been the effect of ionization state of the ISM.
Specifically ionizing radiation from stars may not only produce recombination
nebulae but also heat the diffuse warm medium.  Thus as well as producing \lya,
the further propagation of LyC radiation may increase the ionization levels of
the diffuse gas, lowering the optical depth of the \hII\ regions to Lyman
radiation \citep[e.g.][]{Pellegrini2012}.  Given that the ISM of galaxies may be
very inhomogeneous, and that \lya\ may be absorbed close the nebulae in which it
forms, the propagation of ionizing radiation could potentially have a large
impact upon the first stages of \lya\ transfer.

The ionization parameter (the number of hydrogen-ionizing photons per atom)
governs the excitation of the gas, which is usually quantified observationally
by the excitation parameter (an emission line ratio that contrasts high and low
ionization species).  This is done most effectively by taking ratios of $p^2$
and $p^3$ ions: \citet{Zastrow2011,Zastrow2013} performed `ionization parameter
mapping' of several local starbursts using the [\sIII]$\lambda
9069$/[\sII]$\lambda 6716$ ratio, to identify highly ionized cones that could
signpost LyC emission.  Similar integrated measurements of `Green Pea' galaxies
found high [\oIII]$\lambda 5007$/[\oII]$\lambda 3727$ ratios \citep{Jaskot2013}
that implies high ionization states, and indeed these galaxies have been found
to be high-EW \lya\ emitters (\citealt{Jaskot2014}; Henry et al., in
preparation).

Currently we do not know how these highly ionized regions affect \lya\ on small
scales in the ISM.  Resolved \lya\ imaging has now been obtained for over 50
galaxies with ACS, but as yet the ionization structure has only been mapped in
one of them.  \citet{Bik2015} used VLT/MUSE observations of [\siII] and [\oIII]
to map the ionization parameter and gas kinematics in \eso, and compare with the
\lya\ imaging from \citet{Hayes2005}.  Two outflowing and highly-ionized conical
regions are revealed, that approximately align with the brightest \lya\ regions.
This would be consistent with a scenario in which \lya\ may propagate with less
scattering through these highly ionized regions, but the observations of one
system cannot establish a causal relation, and a large sample of similar
observations needs to be obtained.  Pertinent observational questions include
whether \lya\ emission is systematically enhanced in regions of high
[\sIII]/[\sII], or whether chimneys through the ISM may feed brighter regions of
\lya\ emission in extended halos.

\subsection{Can we Predict \lya\ Observables from Other Information?}

Given the importance of \lya\ observations at both low and high redshift (see
Section~\ref{sect:intro:key}), a vital question becomes whether we can predict
the \lya\ escape fraction, equivalent width, or line profile from a given set of
quantities.  I.e. if we are given a dust content, a characteristic velocity for
outflowing atomic gas, etc, do our predictions for \lya\ emission/absorption
match reality?

Given the absence of strong correlations involving \lya\ (for example in
Figure~\ref{fig:globtrends}), one may be tempted to conclude that our
understanding is not this sophisticated.  However if, on the other hand, the
\lya\ that we observe is governed by geometrical considerations such as viewing
angle, then averaged over many galaxies the answer may be more encouraging.  For
example it has been shown that transfer models inside homogeneous shells can
reproduce very wide ranges of line profiles \citet{Schaerer2011}, similar to
those that are observed globally at high-$z$; furthermore when coupled with
semi-analytical models of galaxy formation, such models are able to reproduce
the broad features of the \lya\ LF over a wide range of redshift
\citep{Garel2012}.  This may imply that our more general picture of \lya\
transport is correct.

It is not clear whether current samples are sufficiently large, or span a high
enough dynamic range in luminosity/SFR for such an exploration.  However in the
coming years, larger databases of global properties will become available; GALEX
LAEs will remain at around 100 objects; HST spectroscopic samples already exceed
this while imaging observations remain somewhat smaller.  This should permit
statistical studies studies using linear discriminant analyses, to determine how
combinations of properties produce the observed \lya\ characteristics.  Many
such possibilities may be envisaged if the signal is strong enough and the
samples are sufficiently large.

More empirically motivated simulations can be performed using the wealth of data
available for low-$z$ galaxies as input.  In such an approach, nature sets up
the ISM instead of computers.  For example, transport calculations have been run
in synthetic galaxies output by hydrodynamical simulations
\citep{Laursen2009,Verhamme2012,Yajima2014}, although as yet there has been no
attempt to construct realistic input conditions of a galaxy based upon
observation.  Such an experiment is not easy, especially without detailed
knowledge of the \hI, but as that becomes available the more the study becomes a
possibility.  With maps of the ionizing stellar population, nebular gas and
ionization structure, \hI\ distribution including large and small scale
kinematics, it will become possible to generate sets of model galaxies that are
based upon real systems for which \lya\ observations have been obtained.
Transfer simulations in such models will then recover the \lya\ morphology and
spectral profile, that can be tested against observation.

\begin{acknowledgements} 
I acknowledge the support of the Swedish Research Council, Vetenskapsr{\aa}det
and the Swedish National Space Board (SNSB).  This research has made use of the
NASA/IPAC Extragalactic Database (NED) which is operated by the Jet Propulsion
Laboratory, California Institute of Technology, under contract with the National
Aeronautics and Space Administration.  I would like to extend warm thanks to  my
friends and collaborators for comments and feedback on the manuscript: John
Cannon, Mark Dijkstra, Daniel Kunth, Peter Laursen, J.~Miguel Mas-Hesse, Jens
Melinder, H\'ector Ot\'i Floranes, Ivana Orlitov\'a, Daniel Schaerer and Anne
Verhamme.  Lucia Guaita is thanked for making equivalent width tables available
for Figure~\ref{fig:lfs}.  Further I would like to thank G{\"o}ran {\"O}stlin,
Claudia Scarlata, Sebastiano Cantalupo, Len Cowie, and many others, especially
those from the 2013 NORDITA \lya\ workshops, for valuable and stimulating
discussions.  I thank the anonymous referee for several careful readings of the
manuscript, and providing many thoughtful comments that have greatly improved
the content. 
\end{acknowledgements}

\bibliographystyle{apj}

\end{document}